\documentclass[twocolumn]{aastex62}
\usepackage{float,graphicx,amsmath,multirow,mathtools,ulem,comment}
\usepackage{color}

\newcommand{\kms}{km~s$^{-1}$ }

\received{Oct 2018}
\revised{Jan 2019}
\accepted{\today}
\submitjournal{ApJ}

\shorttitle{COMs in NGC 1333 IRAS 4A1}
\shortauthors{Dipen et al.}

\begin{document}

%\title{Complex molecules in absorption and the interesting scenarios for NGC1333 IRAS~4A1: Implication of a hot atmosphere/corino from ALMA observation}

\title{ Implications of a hot atmosphere/corino from ALMA observations towards NGC1333 IRAS~4A1}

\author{Dipen sahu}
\altaffiliation{Physical Research Laboratory, Ahmedabad-380009, India.}

\affiliation{ Indian Institute of Astrophysics, Sarjapur Main Road, 2nd Block, Koramangala,Bangalore-560034, India.}
\author{Sheng-Yuan Liu}
\affiliation{Academia Sinica Institute of Astronomy and Astrophysics, 11F of AS/NTU Astronomy-Mathematics Building, No.1, Sec. 4, Roosevelt Rd, Taipei 10617, Taiwan, R.O.C.}
\author{Yu-Nung Su}
\affiliation{Academia Sinica Institute of Astronomy and Astrophysics, 11F of AS/NTU Astronomy-Mathematics Building, No.1, Sec. 4, Roosevelt Rd, Taipei 10617, Taiwan, R.O.C.}
\author{Zhi-Yun Li}
\affiliation{Astronomy Department, University of Virginia, Charlottesville, VA 22904, USA}
\author{Chin-Fei Lee}
\affiliation{Academia Sinica Institute of Astronomy and Astrophysics, 11F of AS/NTU Astronomy-Mathematics Building, No.1, Sec. 4, Roosevelt Rd, Taipei 10617, Taiwan, R.O.C.}
\author{Naomi Hirano}
\affiliation{Academia Sinica Institute of Astronomy and Astrophysics, 11F of AS/NTU Astronomy-Mathematics Building, No.1, Sec. 4, Roosevelt Rd, Taipei 10617, Taiwan, R.O.C.}
\author{Shigehisa Takakuwa}
\affiliation{Department of Physics and Astronomy, Graduate School of Science and Engineering, Kagoshima University, 1-21-35 Korimoto, Kagoshima 890-0065}
\affiliation{Academia Sinica Institute of Astronomy and Astrophysics, 11F of AS/NTU Astronomy-Mathematics Building, No.1, Sec. 4, Roosevelt Rd, Taipei 10617, Taiwan, R.O.C.}

\correspondingauthor{Dipen Sahu}
\email{dipenthink@gmail.com}

\begin{abstract}
%{\bf to be revised after the main paper is finalized}
We report high angular resolution observations of NGC1333 IRAS4A, a protostellar binary including A1 and A2, at 0.84~mm with the Atacama Large Millimeter/submillimeter Array. From the continuum observations, we suggest that the dust emission from the A1 core is optically thick, and A2 is predominantly optically thin. The A2 core, exhibiting a forest of spectral lines including complex molecules, is a well known hot corino as suggested by previous works.
More importantly, we report, for the first time, the solid evidence of complex organic molecules (COMs), including CH$_3$OH, $^{13}$CH$_3$OH, CH$_2$DOH, CH$_3$CHO associated with the A1 core seen in absorption. The absorption features mostly arise from a compact region around the continuum peak position of the A1 core.  Rather than originating from a larger common envelope surrounding the protobinary, the COM features are associated with individual cores A1 and A2. Considering the signatures observed in both continuum and spectral lines, we propose two possible scenarios for IRAS 4A1 - the COM absorption lines may arise from a hot-corino-like atmosphere at the surface of an optically-thick circumstellar disk around A1, or the absorption may arise from different layers of a temperature-stratified dense envelope.

\end{abstract}

\noindent
\section{Introduction} \label{sec:intro}

 Hot cores and hot corinos are associated with (high- and low-mass) star formation at its relatively early stage of evolution. Due to high temperature in the vicinity of protostar, large abundance of complex organic molecules (containing typically at least six atoms, ~\citealp{herbst09}) are observed in the gas phase.
Hot cores, with typical sizes of $<$ 0.1 pc, densities of $> 2\times 10^{7}$ cm$^{-3}$, and temperatures of $>$ 100~K (e.g.,\citealp{vander04}), are related to massive star forming regions.
Orion KL, for example, hosts an archetypal hot molecular core as revealed by millimeter molecular line observations (e.g., \citealp{blake87}).
Similar spectral line surveys found that saturated COMs, e.g., CH$_3$OH, HCOOCH$_3$, CH$_3$OCH$_3$ are especially abundant in the hot core regions (e.g. \citealp{comito05,tercero10,croc14}).
The low mass analogs of the hot cores are dubbed as hot corinos.
NGC 1333~IRAS~4A (IRAS~4A hereafter) and IRAS~16293$-$2422 (I16293 hereafter) are among the first recognized hot corinos \citep{cazaux03, bott04}.
There are only a handful of hot-corinos known to date (e.g.,\citealp{bott07,oberg11,oberg14,fuente14}).

The observed COMs in hot cores and hot corinos have often been suggested to form primarily in grain icy mantles and later get liberated into the gas phase by thermal evaporation (desorption) \citep{millar91,charnley92}. 
Further investigation indicated, though, that their formation through recombination of radicals on grain surfaces during the warm up phase in the star formation process may be as important \citep{garrod06}.
Laboratory experiments and theoretical calculations, nevertheless, demonstrated that methanol is produced largely in the grain phase through repeated hydrogenation of CO (e.g. \citealp{tielens82,tielens05}) as,
\begin{equation}
\rm{CO _{surface} \xrightarrow[]{\text{H}}  HCO_{surface}  \xrightarrow[]{\text{H}}  CH_3O_{surface}
\xrightarrow[]{\text{H}} CH_3OH_{surface}}
\end{equation}
Gas-phase production of CH$_3$OH, on the other hand, is inefficient as compared to grain-phase for typical physical conditions in protostellar environments \citep{geppert06}.
Consequently, methanol is often used to signify the presence of  thermally-evaporated COMs in hot core and hot corinos.

In the context of low-mass YSOs, COMs were found in a few cases specifically associated with young stellar and protoplanetary disks. For example, mythalcyanide (CH$_3$CN) and CH$_3$OH were detected in the protoplanetary disk TW Hya \citep{oberg15,walsh16}.
\citet{lee17b} detected CH$_3$OH, deuterated methanol (CH$_2$DOH), methyl mercaptan (CH$_3$SH), formamide (NH$_2$CHO) as well as doubly deuterated formaldehyde (D$_2$CO) in HH212. The distribution of these molecular emission, above and below the dusty protoplanetary disk, are suggested to trace COMs formed in situ in the atmosphere of the disk. More recently, \citet{vant18} detected methanol that appears to be thermally desorbed from grains in the disk of the young out-bursting source V883 Ori. 

Complex molecular species mainly liberated from grain surfaces through thermal desorption with dust temperature reaching $\sim$85 K (Brown et al. 2007). Additional effects like photodissociation, reactive desorption and energetic interaction may enhance gas-phase abundances of chemical species  (e.g. \citealp{sahu18,taquet15}) including COMs \citep{droz15}.
For this reason, a few COMs have been seen in the peripheral
region of protostellar envelopes (e.g., \citealp{jaber14}) 
and in shocks produced by fast jet and molecular outflow (e.g. \citealp{palau17} and references therein).

Thanks to the advent of the interferometric facilities like the Atacama Large Millimeter and submillimeter Array (ALMA), spectral signatures of COMs can be observed and mapped with high sensitivity and angular resolution at sub-arcsecond level, differentiating their emission regions.
Previously, interferometric observations of hot corinos (e.g.\citealp{bott04,kuan04,bisschop08,jorgen11,jorgen16,codella16}) unveiled the presence of several COMs and in some cases their distributions, too.

The IRAS~4A, one of the first known hot corino objects, is located in the Perseus molecular cloud at a distance 235~pc \citep{hiro08}. 
A recent result from GAIA measured the distance of Perseus molecular cloud to be 293$\pm$22~pc \citep{ortiz18, zucker18}, we adopt this updated distance ($\sim$293 pc) throughout the paper.
%However, throughout the text, we consider the distance to be 235~pc for easy comparison with earlier results.  
The overall luminosity and envelope mass of the object are 9.1 L$_\odot$ and 5.6 M$_\odot$ respectively \citep{kris12,karska13}. 
In the centimeter to millimeter continuum imaging measurements, IRAS~4A shows two compact emission cores and clumpy extended structure, and IRAS~4A1 (A1 hereafter) is found to be brighter than IRAS~4A2 (A2 hereafter) (\citealp{choi10,choi11}).
A1 and A2, separated by 1\farcs8  ($\sim$ 527~AU), are likely a pair of class 0 proto-binary system originating from the same parent cloud \citep{jenn87,sand91,lay95,looney00} and are reported to have similarly powerful bipolar outflows \citep{santg15}.
The IRAS~4A region therefore encompasses circumstellar disks associated with compact cores, and protostellar envelopes around the individual protostars and a common envelope shared by the binary \citep{choi11}.
\citet{taquet15}, through comparatively low angular resolution ($\sim 2''$) observations that were unable to disentangle A1 and A2, found the presence of several COMs, e.g., CH$_3$OH, HCOOCH$_3$, CH$_3$OCH$_3$, HCOCH$_2$OH, C$_2$H$_5$OH towards IRAS~4A as a whole.
Recently, \citet{lopez17} studied the complex organics in IRAS~4A with ALMA and the Pleatu de Bure Interferometer (PdBI) at a higher angular resolution ($\sim 0.5''$) and resolved apart the emissions from the two cores, A1 and A2.
They found a striking contrast between these two neighboring cores ---
while A2 showed hot corino activity with enriched COMs' emission \citep{persson12}, no sign of COMs was detected in A1. 
\citet{lopez17} suggest that either A1 does not host any hot corino, or alternatively A1 may host a hot corino with a size  $\sim$15 AU (after scaling the adopted distance to 293 pc) that is 6 times smaller than that of  A2.

Keeping all this in mind, in this paper we present ALMA observations of the protostellar binary IRAS~4A with a high resolution of $\sim 0.3''$ (or equivalently 88 AU in linear scale). We detected emission signatures of methanol and its various isotopologues as well as CH$_3$CHO in A2, while these species showed primarily absorption features towards A1.
We suggest that we are seeing COMs in the ``atmosphere" of A1, 
which itself is readily developing into a hot corino object with optically thick dust continuum emission.
We describe the observational setup in Section 2, detail the results in Section 3, discuss the implications of these results in Section 4.

\section{Observations} \label{sec:style}

The observations of IRAS~4A were carried out by ALMA under the project code 2015.1.00147.S. 
Three Execution Blocks (EBs) were conducted with ALMA Band 7 receiver on 2016 Jul. 23 and 24 and 2016 Dec. 14. 
The number of antennas in the 12~M array for the three EBs were respectively 39, 42, and 43.
The phase centre of the observations was set at R.A.(J2000) = 03$^d$ 29$^m$ 10.50$^s$ , Dec.(J2000) =+31$^\circ$13$' $ 31.5$''$ and the total integration time on-source was approximately 84~mins.
We deployed seven spectral windows, including one broadband window centered at 350.714 GHz whose bandwidth and spectral channel width are 1.875 GHz and 976 kHz, respectively.
The latter, with standard online Hanning-smooth applied, corresponds to a velocity resolution of 0.84 km s$^{-1}$.
This paper focuses on the data from the above broadband window and more details about the full observational setup can be found in our forthcoming paper (Su et al. 2019)
%\citet{su18}.

The data were first calibrated by the ALMA observatory through the data reduction pipeline within the Common Astronomy Software Application package (CASA, version 4.7 \citealp{mcmullin07}), 
We then generated both the 0.84~mm continuum and spectral visibilities by fitting and subtracting continuum emission in the visibility domain.
We used Briggs weighting with a robustness parameter of 0.5 for forming the images. 
The resulting synthesized beam size is 0\farcs30 $\times$ 0\farcs20 (PA=-6.45$^\circ$) for the continuum map and is 0\farcs31 $\times$ 0\farcs20 (P.A= -24$^\circ$) for the spectral data cubes (for a typical line transition only).
We subsequently used MIRIAD \citep{sault95} and CASSIS (developed by IRAP-UPS/CNRS- http://cassis.irap.omp.eu) for further image inspection and spectral line analysis.
The resulting rms level is $\sim$ 8 mJy~beam$^{-1}$ and 3 mJy~beam$^{-1}$ in the continuum image and spectral cubes, respectively, with the noise rms being strongly limited by the imaging dynamical range in the continuum image due to the presence of very bright features.

\section{Results} \label{results}

\subsection{Dust continuum emission \label{sec:res-dust}} 

Figure~\ref{spectral_maps}(a) shows the 0.84~mm ($\sim$357 GHz) continuum towards IRAS~4A.
At an angular resolution of 0\farcs31 $\times$ 0\farcs20, IRAS~4A is clearly resolved into two components, with A1 located in the SE, A2 in the NW.
A1 and A2 each appears to consist of a compact emission feature and an outer extended feature. We applied a two-component 2D Gaussian fitting to the two sources and summarize the fitting results in Table~\ref{tb-cont-fit}. Overall we recovered within the region a flux density of 7.24 Jy.
Previous single dish observations by \citet{smith00} at 0.85~mm reported a peak flux density of 10.3 Jy~beam$^{-1}$ within a 16${''}$ beam toward IRAS~4A. \citet{sand01}, on the other hand, measured at 0.85~mm a peak flux density of 9.05 Jy towards the same direction. Considering calibration uncertainties, these results appear consistent and are about only 20$\%$ larger than our ALMA measurement.
Given the integrated region of single dish observations is also larger than the extent of the IRAS~4A centroid, we conclude that the missing flux issue, though may result from the most extended envelope, is not severe and does not impact our science focusing on the compact continuum features.

The angular separation between A1 and A2 derived from the positions of the compact components listed in Table~\ref{tb-cont-fit} is 1\farcs8, in agreement with previous measurements \citep[e.g.,][and references therein]{lopez17}.
For both A1 and A2, the diameters of the compact component are equivalent to about  55 -- 115 AU, and the sizes of the extended envelope span about several hundreds to nearly 1000 AU.
The brightness temperatures derived from the observed peak continuum are 57 K and 42 K towards A1 and A2, respectively. 
The relatively high brightness temperatures readily indicate that the center positions of both A1 and A2 are very warm. We note that these brightness temperatures are the beam-average values, and the actual brightness temperatures, depending on the source filling factor, could be further inflated.

Assuming a gas-to-dust ratio of 100, a (uniform) dust temperature of 60 K, a dust opacity $\kappa_{\nu}$ = 0.006 ($\nu$/245 GHz) $^\beta$ cm$^2$ g$^{-1}$ (Kramer et al. 1998; Shepherd \& Watson 2002), an opacity index $\beta$ = 1.5, and a distance of 293 pc to IRAS 4A, the dust and gas mass is estimated to be  0.15, 0.49, 0.07, and 0.50 \emph{M}$_\odot$ for 4A1 compact component, 4A1 extended component, 4A2 compact component, and 4A2 extended component, respectively; calculations are based on optically thin assumption. 
Using the above-mentioned parameters, the H$_2$ column density toward the 4A1 centroid is estimated to be at least 1.3 $\times$ 10$^{26}$ cm$^{-2}$. Such a high molecular gas column density will lead to a visual extinction A$_v$ $>$100000.  
Note that the estimated dust and gas mass of the 4A1 compact component is most likely a low limit given that there are indications of the A1 dust continuum being optically thick throughout the radio to submillimeter bands as discussed in Section~\ref{sec:dis-cont}.

\begin{deluxetable*}{llccrccrc}
%\label{tb-cont-fit}
\singlespace
 \tabletypesize{\footnotesize}
 \tablecaption{\sc Parameters of Continuum Sources \label{tb-cont-fit}}
 \tablewidth{0pt}
   \tablehead{\colhead {Source} & \colhead{Component}  & \multicolumn{2}{c}{Position (ICRS 2000)} & \colhead{S$_{int}$} &\multicolumn{3}{c}{Deconvolved Size} & \colhead{Mass} \\
            \colhead{} & \colhead{} & \colhead{$\alpha$} & \colhead{$\delta$} & \colhead{(mJy)} & \colhead{max$\arcsec$} & \colhead{min$\arcsec$} & \colhead{P.A.$^\circ$} & \colhead{\emph{M}$_\odot$} }
 \startdata
        4A1	& compact	& 03 29 10.538  & +31 13 30.93	& 923 	& 0.397 & 0.354 & 64.1 	& 0.15 \\
		& extended	& 03 29 10.523	 & +31 13 30.69	& 2979	& 1.937 & 1.282 & 47.0 	& 0.49 \\
        4A2  & compact       	& 03 29 10.430  & +31 13 32.08	& 404	& 0.274 & 0.168 & $-$82.2 & 0.07 \\
		& extended       	& 03 29 10.448  & +31 13 32.12	& 3035 	& 2.973 & 1.167 & $-$53.8 & 0.50\\ 
\enddata
%\tablenotetext{a}{at peak of CO (2$-$1)}
%\label{tb-cont-fit}
\end{deluxetable*}

\subsection{Detected molecules and their
distribution \label{sec:res-line}}

In Figure~\ref{line_spectra} we display the spectra towards the continuum A1 and A2 peaks over the full 1.875~GHz spectral window centered at 350.714~GHz.
Line features from methanol (CH$_3$OH) and its isotopologues ($^{13}$CH$_3$OH and CH$_2$DOH) as well as those from acetaldehyde (CH$_3$CHO), as identified and discussed below, are marked in Figure~\ref{line_spectra} and listed in more detail in Table~\ref{line_list}.

Towards A1, the spectrum throughout the window exhibits rich and almost exclusively absorption features, regardless of their excitation temperatures.
This is a complementary evidence suggesting that the continuum dust emission along the A1 line of sight is optically thick.
In contrast, the spectrum towards A2 shows a multitude of emission lines. 
Many spectral features (e.g., CH$_3$OH) from A2 display noticeably similar peak intensities at around 75~K, indicative of a ``saturated" and thus optically-thick nature of those respective lines. 

In this paper, we focus on the detection and analysis of CH$_3$OH, its isotopologues including $^{13}$CH$_3$OH and CH$_2$DOH, and CH$_3$CHO in both A1 and A2 cores of IRAS~4A.
As noted, CH$_3$OH is a robust tracer for highlighting surface chemistry which occurs within interstellar ices on the grain surface.
To firmly secure the spectral identification, we examined the spectral features that bear strong (emission/absorption) intensities and are relatively well separated from line confusion/contamination.
We visualized the spectral data in CASSIS for initial line identifications.
Assuming a systematic velocity 6.96 \kms \citep{difran01} and
employing the JPL \citep{pick98} and CDMS \citep{muller01,muller05} databases, we generated synthetic spectra within CASSIS.
Essentially all isolated transitions from these molecules with their Einstein co-efficient $A_{ij} \geq 10^{-5}$ are well matched (at a level of 3$\sigma$ and above) while those with $A_{ij} < 10^{-5}$ are likely below our detection threshold or blended by nearby stronger features from other species. 
CH$_3$OH transitions are most intense and there is negligible contamination from other molecular species.
All CH$_3$OH transitions falling in the spectral coverage are identified, as listed in Table~\ref{line_list}.
In particular, the transition 1(1,1)-0(,0,0)++ to the ground state are prominently detected with a deep absorption feature toward A1 and an inverse P-Cygni profile towards A2.
%Similarly, $^{13}$CH$_3$OH, CH$_2$DOH, CH$_3$CHO lines have been detected without line confusion.
Though CH$_2$DOH molecular transitions are detected without confusion, some of its transitions are blended themselves.
For example, two CH$_2$DOH transitions (8$_{(4,4)} - 8_{(3,6)}~e1$ \& 8$_{(4,5)} - 8_{(3,5)}~e1$),
with similar excitation energies E$_u$ and frequencies, overlap at around 349.864 GHz.
The detected molecular transitions of the selected species are common towards both the cores, except that the molecular transitions with E$_u ~ > 240.5$~K of CH$_3$CHO were not detected towards A1. 

We applied Gaussian profile fitting for extracting the spectral parameters of the identified features. 
We list the fitting results in the Table~\ref{line_list} and plot the profiles in Figure~\ref{A2_spectralfit} and Figure~\ref{A1_spectralfit}. 
During the fitting process, we consider the velocity range carefully to fit the spectral profile. 
In some cases, we excluded some channels to avoid line contamination and other kinematics feature (e.g. inverse P-Cygni, see Fig.~\ref{pcygni}). 
The fitting results of spectral profiles from different transitions of a molecule have peak around a common LSR velocity, considering the uncertainties of measurements.
The averaged velocity of the emission features (excluding inverse P-Cygni profile) toward A2 is 6.6 km s$^{-1}$.
Given the limited velocity resolution, this velocity is consistent with the gas velocity seen in previous works (e.g., average value 6.8  km s$^{-1}$, \citealp{lopez17}).
The averaged velocity of absorption features toward A1 is 6.8 km s$^{-1}$. 
Meanwhile, there does not appear any trend of velocity variation among the transitions of different excitation energy.
When this is compared with the systemic velocity of 6.7 km~s$^{-1}$ adopted by \citet{choi07},  there is no clear indication of (infall) motion in the absorbing gas.
Based on the profile fitting (Table 2), we also note that the line-widths $\Delta V$s are $\sim$ 1.0-1.5 km s$^{-1}$ for the (absorption) features in A1 while $\Delta V$s are $\sim$ 1.5-2.3 km s$^{-1}$ for the (emission) lines in A2 core.

\begin{deluxetable*}{llllllllllllll} 
\rotate
\tabletypesize{\scriptsize}
%\tablewidth{0pc}
\tablecaption{Molecular transitions detected towards IRAS~4A\label{line_list}} 
%\tablehead{\colhead{Parameter}                      & \colhead{249.06 GHz}        & \colhead{287.05 GHz}        & \colhead{303.01 GHz}      & \colhead{344.05 GHz}}  
\tablehead{
 \colhead {Molecule} &\colhead{ Transition$^*$} &\colhead { $\nu_0$ (MHz)} & 
 \colhead { E$_{up}$ (K)} & \colhead { A$_{ij}$ sec$^{-1}$} & \colhead {Intensity(K)} & \colhead {V$_{lsr}$ (km~s$^{-1}$)} &\colhead{$\Delta$ V}& \colhead {Intensity(K)} & \colhead {V$_{lsr}$ (km~s$^{-1}$)} &\colhead{$\Delta$ V}
 }
 \startdata
 \colhead{} &\colhead{}&\colhead{}&\colhead{}&\colhead{}&\multicolumn{3}{c}{transitions towards IRAS 4A2}& \multicolumn{3}{c}{transitions towards IRAS 4A1$^{**}$}\\
\hline
${\rm CH_3OH}$       & vt=0 4(0,4) - 3(-1,3) &350687.73 &36.33 & 8.67E-5&  31.05 $\pm$3.06& 6.30 $\pm$0.13& 1.93 $\pm$0.25    &  -7.69 $\pm$0.80& 7.03 $\pm$0.08& 1.25 $\pm$0.16\\
${\rm CH_3OH}$& vt=0	1(1,1) - 0(0,0)++   &350905.12 &16.84 & 3.31E-4 &  38.49 $\pm$2.72& 6.59 $\pm$0.10& 2.06 $\pm$0.12  &  -20.59 $\pm$5.95& 7.10 $\pm$0.24& 1.18 $\pm$0.38\\ 
${\rm CH_3OH}$& vt=0  9(5,5) - 10(4,6)      & 351236.34 &240.51 &3.66E-5 &  27.76 $\pm$1.47& 6.49 $\pm$0.07& 2.09 $\pm$0.14   &  -3.00$\pm$0.15& 6.64 $\pm$0.03& 0.96 $\pm$0.06\\
\hline
 ${\rm ^{13}CH_3OH}$ &vt=0 1(1,0) - 0(0,0)++ & 350103.12 &16.80& 3.29E-4 &  25.59 $\pm$0.86& 6.65 $\pm$0.03& 1.49 $\pm$0.06 & -2.19 $\pm$0.17& 7.01 $\pm$0.05& 1.06 $\pm$0.10\\
 ${\rm ^{13}CH_3OH}$ &vt=0 8(1,7) - 7(2,5)& 350421.58 &102.62& 7.03E-5 &  22.91 $\pm$4.58& 6.53 $\pm$0.29& 1.60 $\pm$0.66 &  -1.52 $\pm$0.18& 6.58 $\pm$0.08& 1.13 $\pm$0.18 \\  
 \hline
CH$_2$DOH & 8(4,4) - 8(3,6) ~e$_1$ & 349864.35 &	149.21&  1.07E-4 & 28.33 $\pm$1.06& 6.65 $\pm$0.09 & 1.54 $\pm$0.14 & -1.84 $\pm$0.29 & 6.76 $\pm$0.12 & 1.05 $\pm$0.29 \\
CH$_2$DOH & 8(4,5) - 8(3,5) ~e$_1$ & 349862.11 &149.21 & 1.07E-4     & 25.23 $\pm$0.13 & 7.00 $\pm$0.02 & 1.65 $\pm$0.02 & -2.60 $\pm$0.41& 6.82 $\pm$0.16 & 1.02 $\pm$0.21 \\
CH$_2$DOH &7(4,4) - 7(3,4) ~e$_1$ \tablenotemark{a} & 349951.68 & 132.07 & 1.00E-4 & 
   30.93 $\pm$1.29& 6.15 $\pm$0.06& 2.14 $\pm$0.13 & -2.69 $\pm$0.43& 6.60 $\pm$0.14& 1.29 $\pm$0.24\\
CH$_2$DOH &  6(4,3) - 6(3,3) ~e$_1$ \tablenotemark{b}& 350027.35 & 117.08& 9.06E-5 & 
 30.86 $\pm$0.87& 6.38 $\pm$0.03& 1.86 $\pm$0.07 & -3.10 $\pm$0.43& 6.61 $\pm$0.08& 1.01 $\pm$0.18 \\
CH$_2$DOH &  5(4,2) - 5(3,2) ~e$_1$ \tablenotemark{c} & 350090.24 & 104.23& 7.58E-5 &
 30.10 $\pm$1.29& 6.38 $\pm$0.05& 1.83 $\pm$0.10 & -2.25 $\pm$0.14& 6.61 $\pm$0.06& 1.33 $\pm$0.10\\
CH$_2$DOH &  4(4,1) - 4(3,1) ~e$_1$ \tablenotemark{d} &350141.30 & 93.53& 5.03E-5   & 
 26.73 $\pm$2.16& 6.38 $\pm$0.11& 2.16 $\pm$0.21 & -2.25 $\pm$0.26& 6.60 $\pm$0.09& 1.20 $\pm$0.16 \\
CH$_2$DOH &  6(2,5) - 5(1,5) ~e$_1$ &350453.86& 71.55& 1.39E-4 & 
27.29 $\pm$1.35& 6.79 $\pm$0.06& 1.75 $\pm$0.11 & -2.23 $\pm$0.08& 7.02 $\pm$0.03& 1.06 $\pm$0.04\\
CH$_2$DOH & 5(1,4) - 5(0,5) ~e$_1$ & 350632.07 & 48.98 & 2.07E-4 &
28.98 $\pm$2.07& 7.05 $\pm$0.09& 1.90 $\pm$0.19 & -2.39 $\pm$0.30& 7.30 $\pm$0.10& 1.18 $\pm$0.17\\     
\hline
CH$_3$CHO&v=0 18(3,15) - 17(3,14) E $^ b$ & 350133.42&179.2&1.44E-3&
28.17 $\pm$3.12& 6.03 $\pm$0.22& 2.30 $\pm$0.45 & -3.58 $\pm$0.31& 6.30 $\pm$0.06& 1.18 $\pm$0.13\\ 
CH$_3$CHO&v=0  18(1,17) - 17(1,16)~E     & 350362.84 &163.46&1.47E-3 &
   26.16 $\pm$1.28& 6.21 $\pm$0.08& 2.23 $\pm$0.17 & -1.84 $\pm$0.19& 6.36 $\pm$0.11& 1.60 $\pm$0.20\\
CH$_3$CHO&v=0  18(1,17) - 17(1,16) ~E      &350445.78 &163.42&1.47E-3& 
26.51 $\pm$2.02& 6.66 $\pm$0.10& 2.00 $\pm$0.20 & -2.70 $\pm$0.04& 6.96 $\pm$0.01& 1.23 $\pm$0.02 \\
CH$_3$CHO&  18(3,15)3 - 17(3,14)3   & 350940.56 &383.69&1.45E-3& 15.15 $\pm$0.71& 6.79 $\pm$0.05& 1.63 $\pm$0.09 & --   &  --   &  --  \\
CH$_3$CHO& 18(1,17)5 - 17(1,16)5     & 351118.83 &368.32&1.45E-3& 16.04 $\pm$0.43& 6.71 $\pm$0.03& 1.60 $\pm$0.06 & --  &  --   &   --  \\
\hline
\enddata
\tablenotetext{a}{4 transitions}\tablenotetext{b}{2 transitions.}\tablenotetext{c}{2 transitions.}\tablenotetext{d}{2 transitions.}
\tablenotetext{*}{Transitions are in the form of N (K$_a$,K$_b$) p v and
N (K$_a$,K$_b$) v.} 
\tablenotetext{**} {The value of intensities are negative as transitions are detected in absorption.}
\end{deluxetable*}

\begin{figure*}[ht]
%\hskip -3cm
\includegraphics[width=18.5cm]{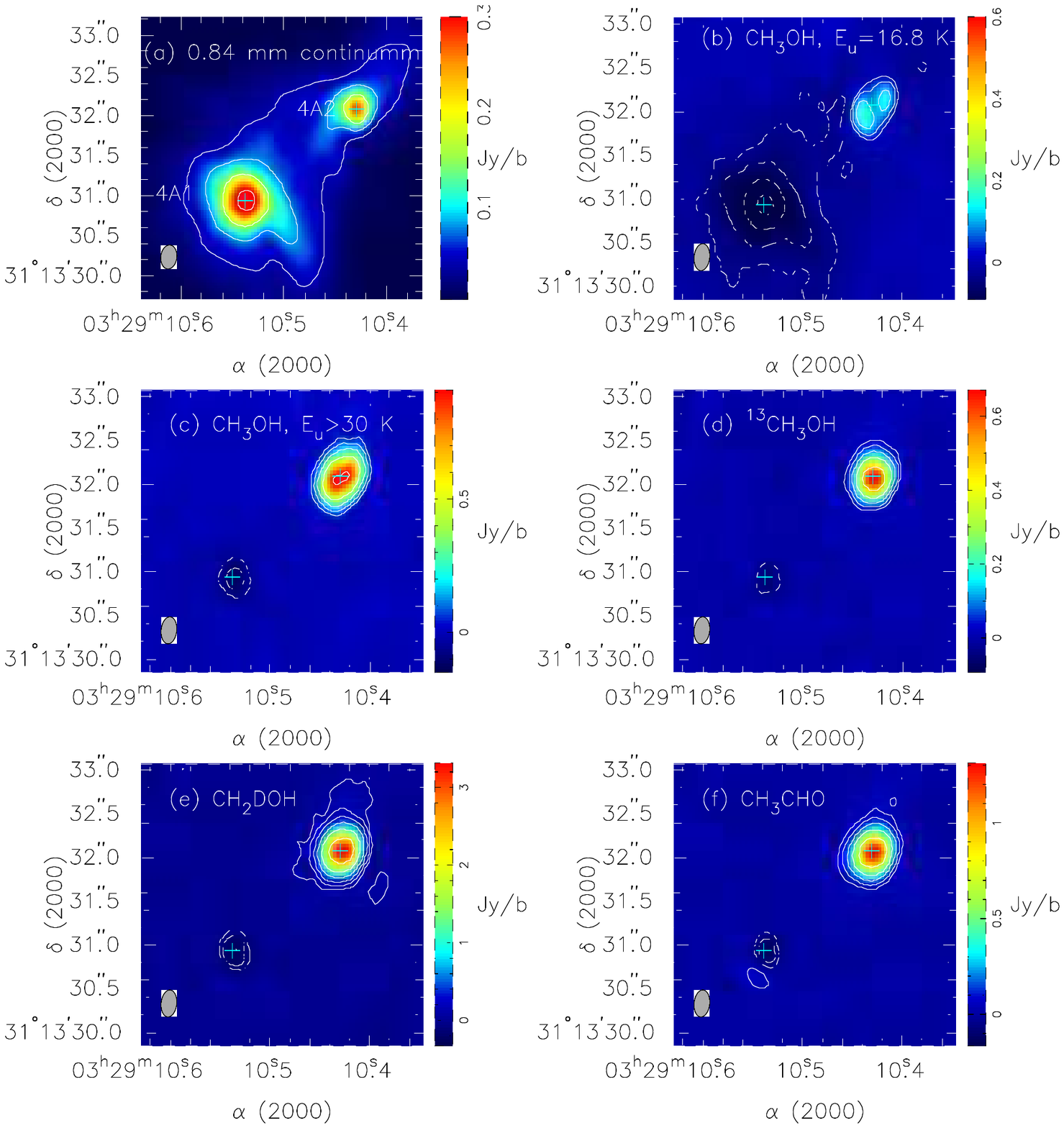}
\caption{%{\bf label (a)-(f) at the top-left concern in front of the continuum/molecule} 
Figures shows integration map of IRAS 4A system for continuum (panel a) and different molecules (b-f) that are discussed in text .
Panel (b) is for 1$_{(1,1)}-0_{(0,0)++}$ transition, and panel (c) is for 4$_{(0,4)}-3_{(-1,3)}$
\& 9$_{(5,5)}-10_{(4,6)}$ transition of CH$_3$OH. Rest of the panel d,e,f are for the molecules $^{13}$CH$_3$OH, CH$_3$CHO,
CH$_2$DOH respectively.  For, the panel d, e,f we consider all the transitions of respective molecules which are detected in both A1 and A2. In the integrated maps we have consider a total velocity interval 2.4 \kms. The contours 
are for 5$\sigma$, $2 \times 5 \sigma$, $2 \times 10 \sigma$ ...., and -5$\sigma$, $2 \times 5 \sigma$, $2 \times 10 \sigma$ ..., with $\sigma$=8,6,11,6,15,10 mJy km s$^{-1}$ for images in panel a,b,c,d,e,f respectively.}
\label{spectral_maps}
\end{figure*}

\begin{figure*}[ht]
\includegraphics[width=18cm]{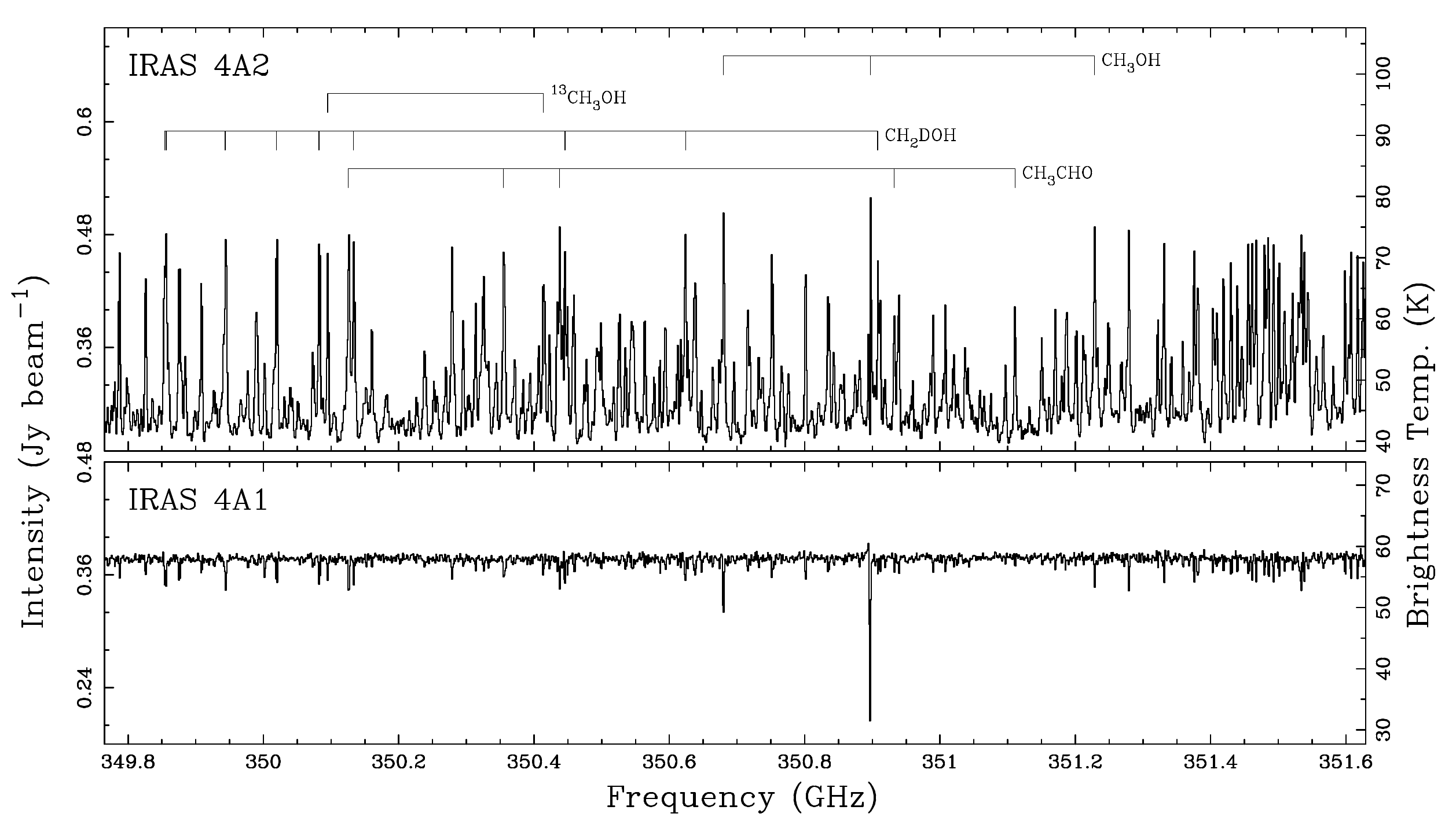}
\caption{Upper panel shows spectral emission towards A2. Spectral transitions towards another core of IRAS 4A, A1 are detected in absorption (lower-panel). Some of the identified transitions for the molecules CH$_3$OH, $^{13}$CH$_3$OH, CH$_3$CHO,
CH$_2$DOH are marked. Along the Y-axis we show real brightness temperature and intensity considering the continuum emission. }
\label{line_spectra}
\end{figure*}

\begin{figure*}[ht]
\centering
\includegraphics[width=13.5cm]{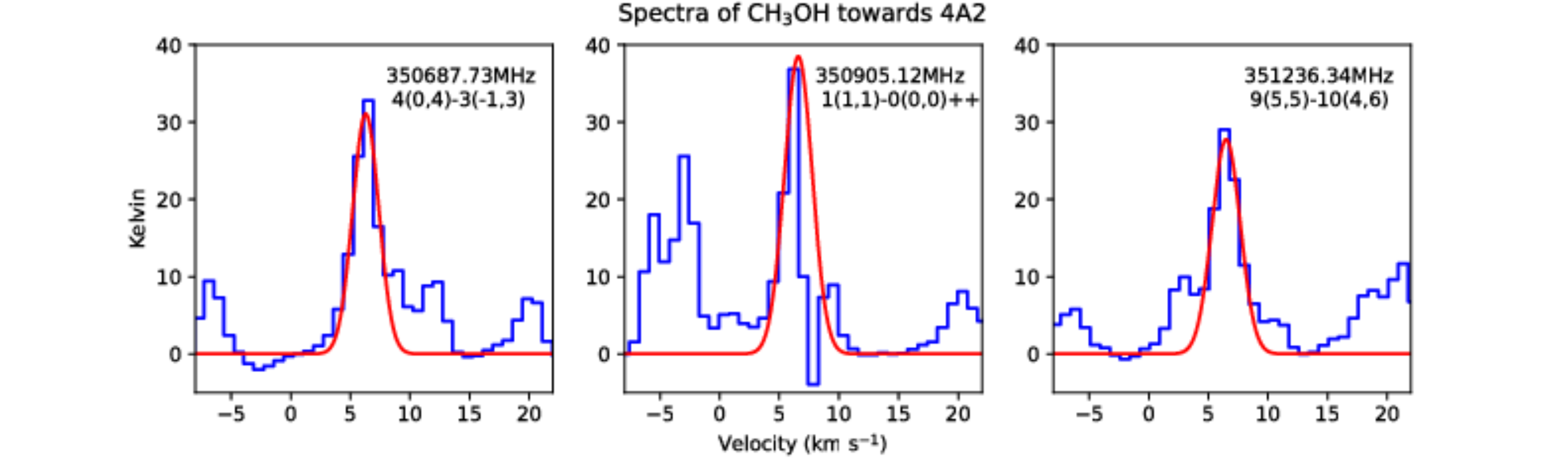}
%\vskip -0.1cm
\includegraphics[width=13.5cm]{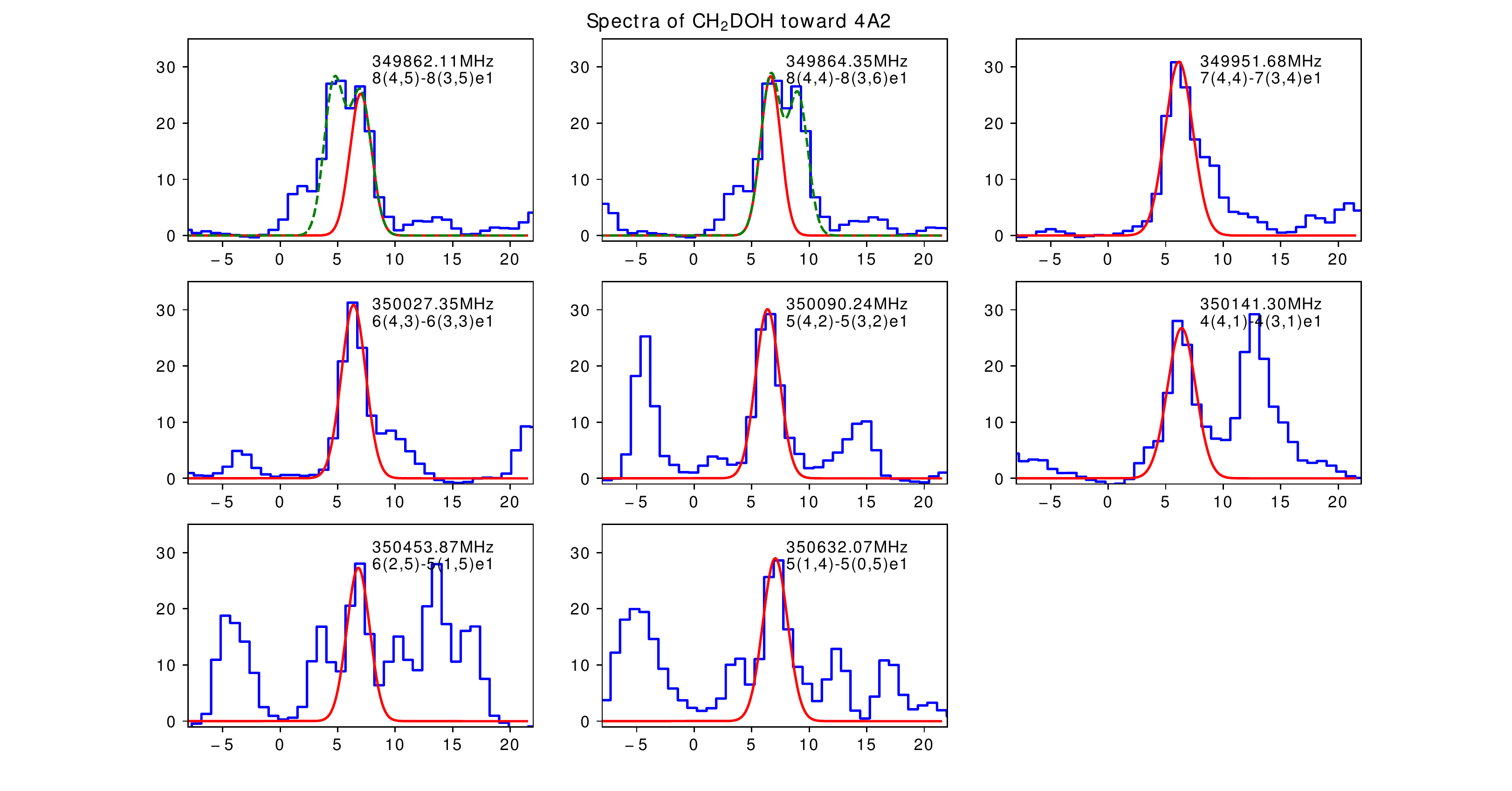}
\includegraphics[width=13.5cm]{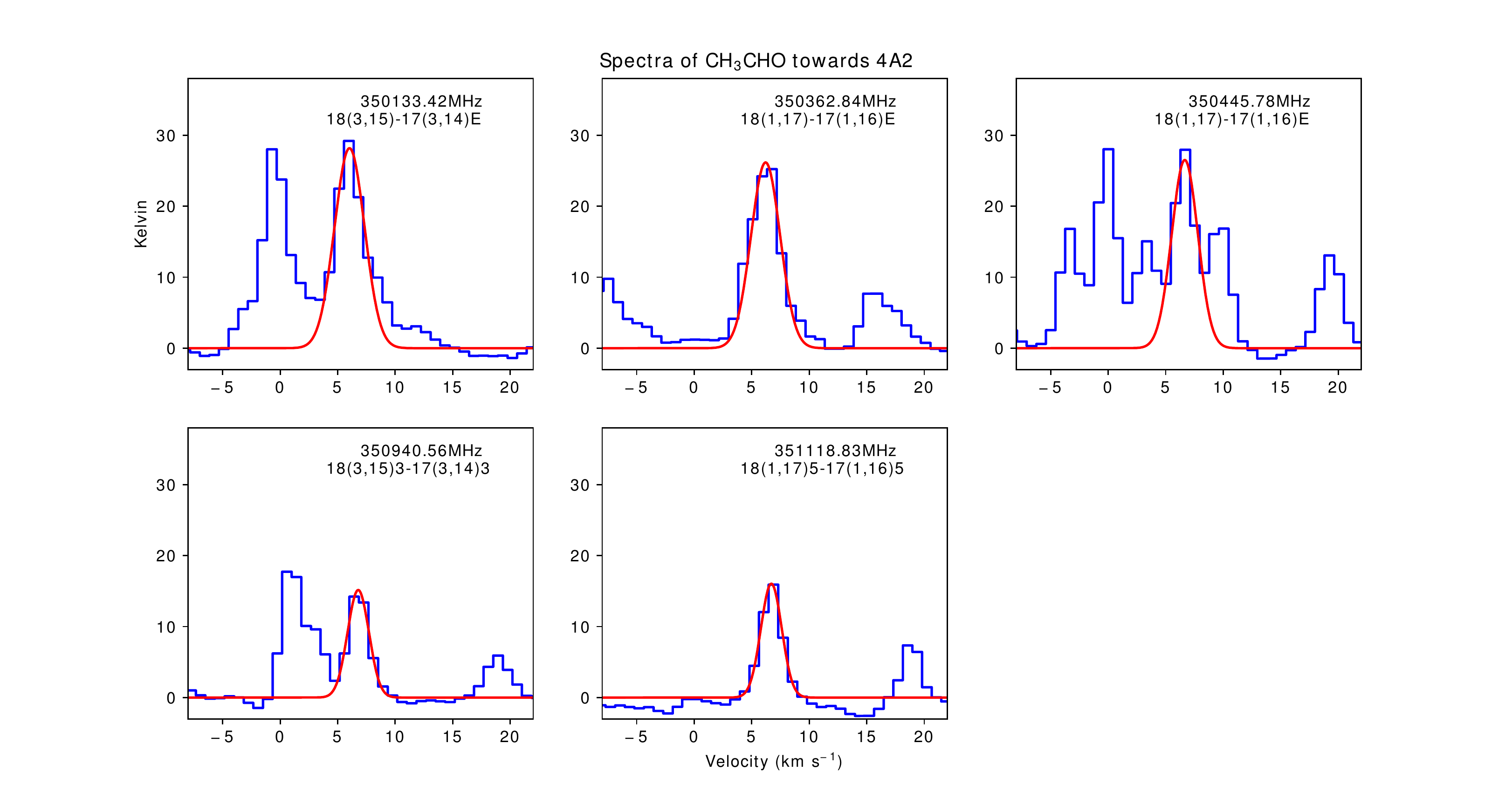}
\vskip -0.2cm 
\includegraphics[width=13.5cm]{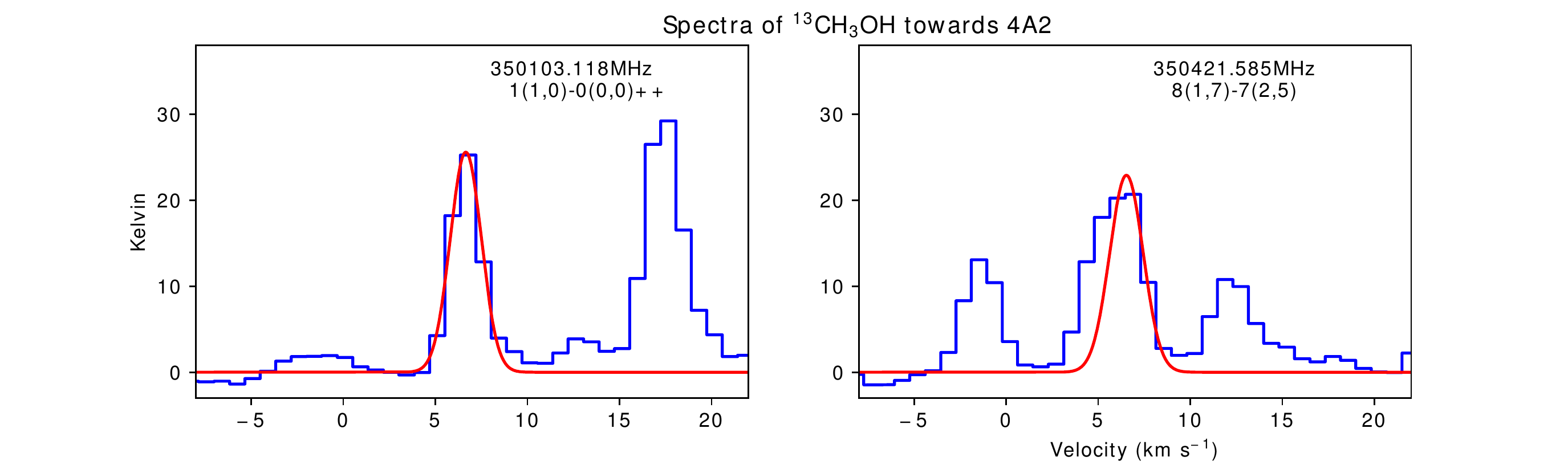}
\caption{Spectra of 18 lines (blue) are detected towards IRAS 4A2. Gaussian fitting of the spectra are shown in red, results of the fittings are given in Table~\ref{line_list}. Here , we show the continuum subtracted intensity of the spectra.}
\label{A2_spectralfit}
\end{figure*}

\begin{figure*}[ht]
\centering
\includegraphics[width=8cm]{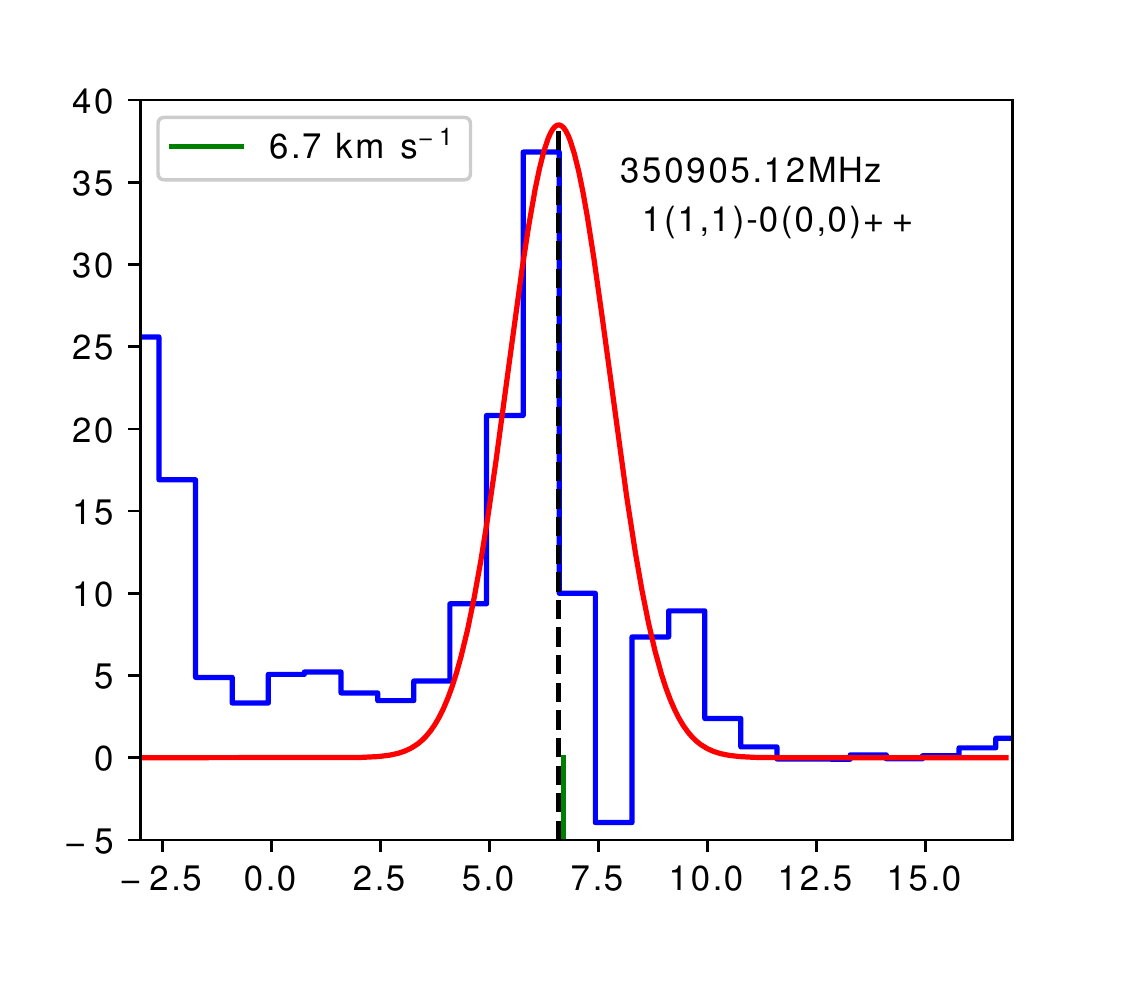}
\caption{The zoomed in view of inverse P-Cygni profile of CH$_3$OH transition towards A2. The dotted line represents the peak velocity of Gaussian fitting, assumed systematic velocity is 6.7 \kms .}
\label{pcygni}
\end{figure*}

\begin{figure*}[ht]
\centering
\includegraphics[width=13.5cm]{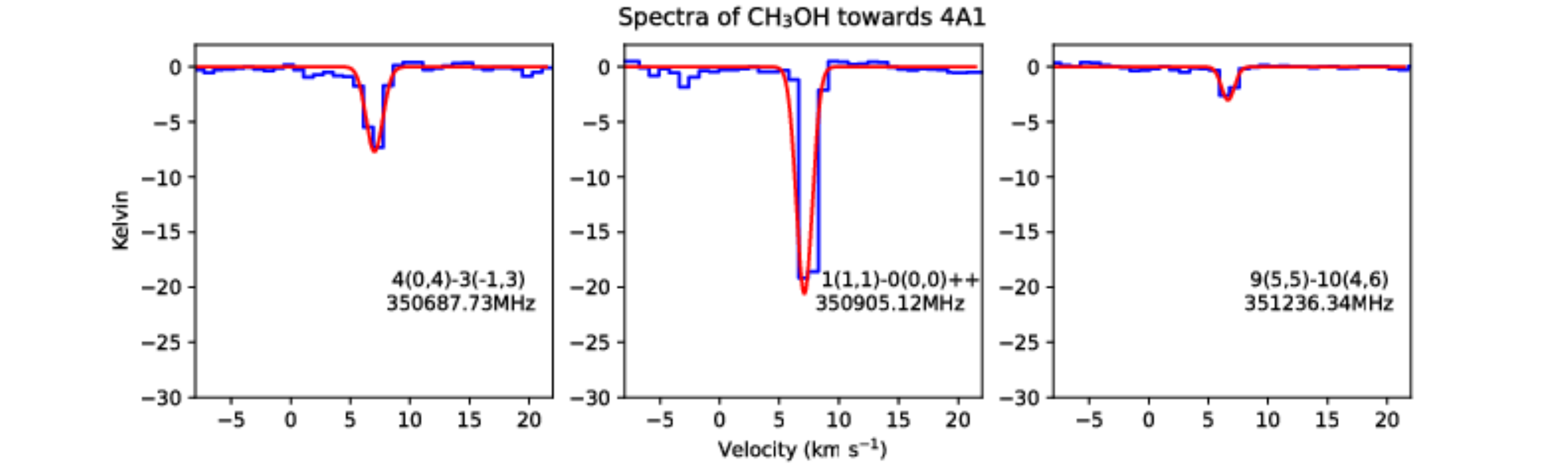}
%\vskip -0.05cm
\includegraphics[width=13.5cm]{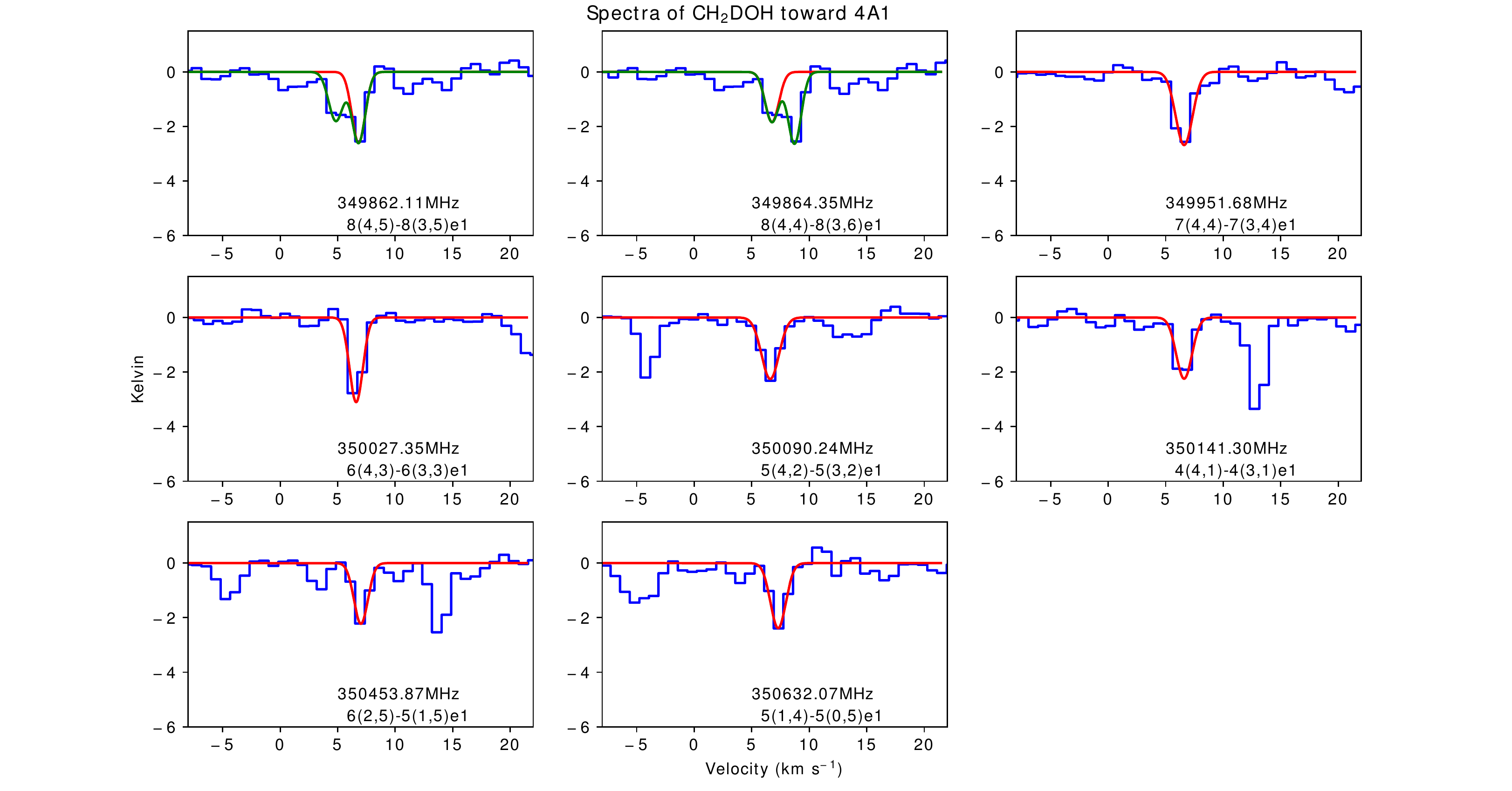}
\includegraphics[width=13.5cm]{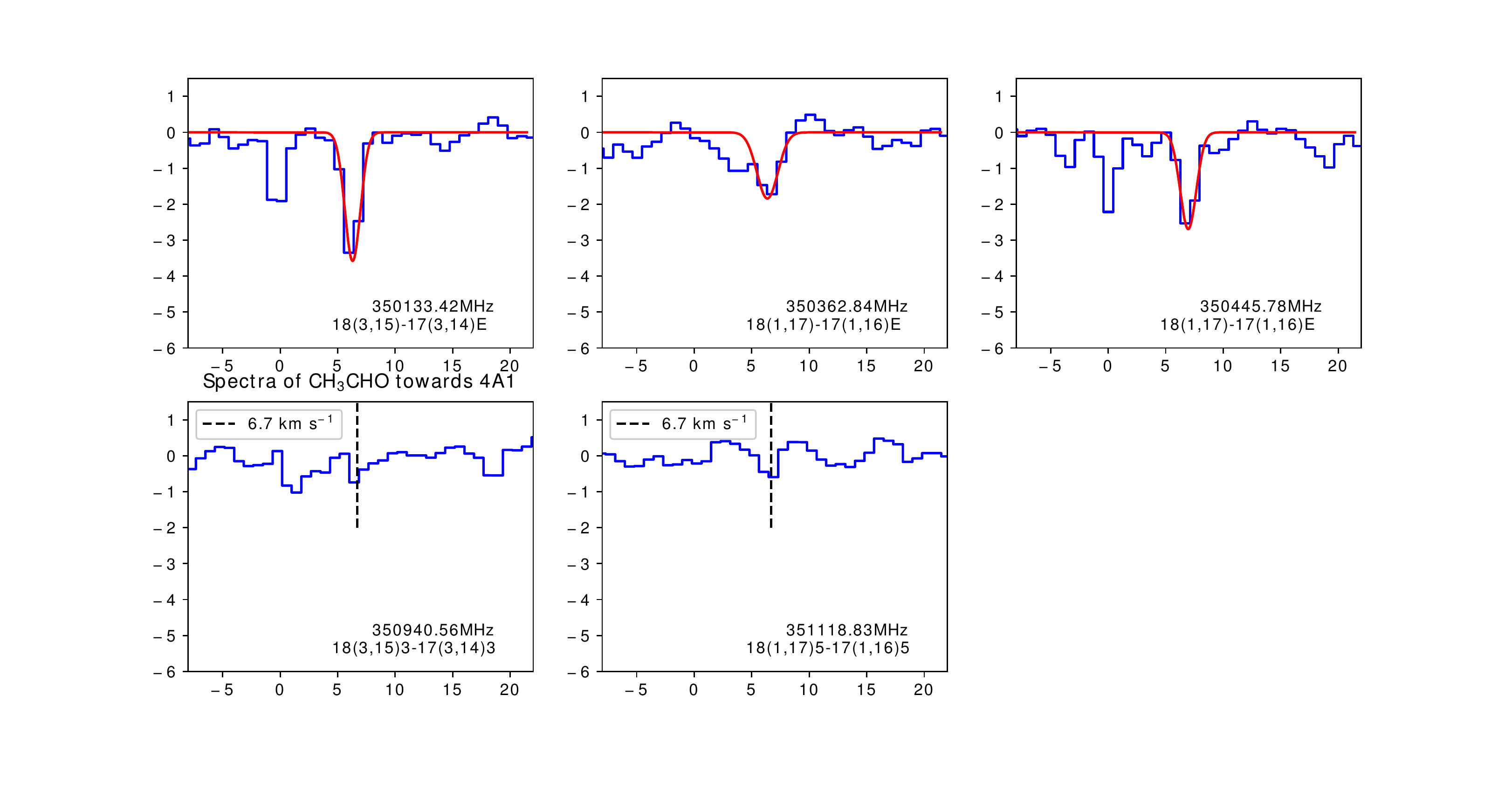}
\includegraphics[width=13.5cm]{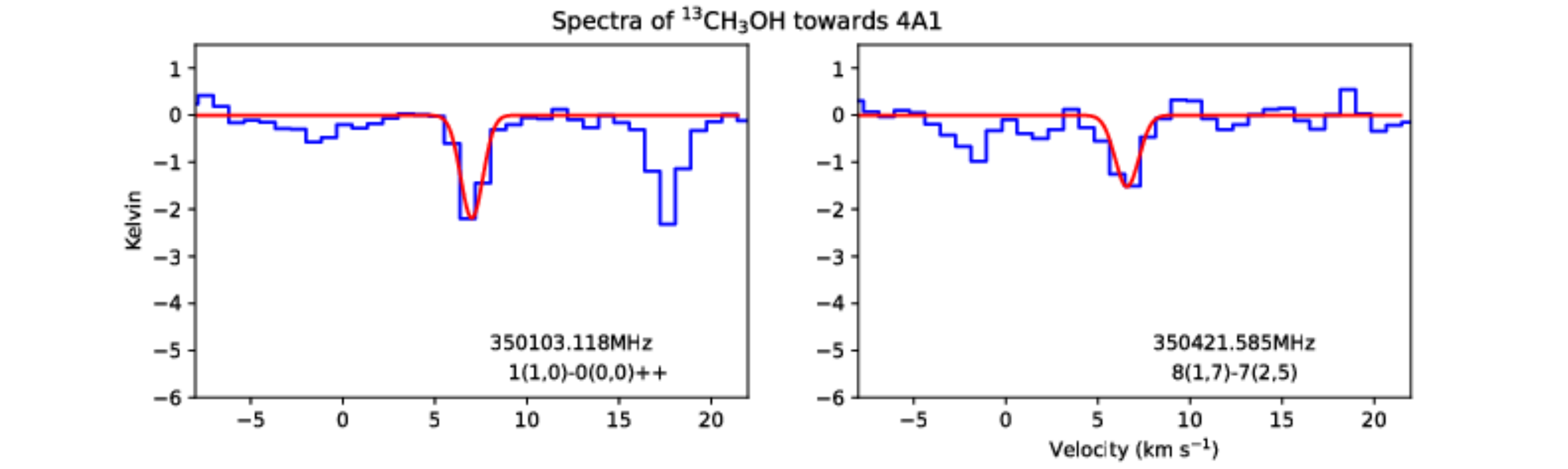}
\caption{Spectra of 16 lines (blue) are detected in absorption towards IRAS 4A1. Gaussian fitting of the spectra are shown in red, results of the fittings are given in Table 2. Here , we show the continuum subtracted intensity of the spectra.}
\label{A1_spectralfit}
\end{figure*}

Figure~\ref{spectral_maps}(b-f) show the velocity integrated intensity maps of the molecules described in the text.
In all panels, dashed contours in the map represent absorption and solid contours represent emission. 
There are three CH$_3$OH transitions detected in both A1 and A2.
We plotted the integrated emission of the low-temperature E$_u$=16.8 K transition, and the remaining two transitions (see Table 2) separately in Fig~\ref{spectral_maps}(b) \& (c).
From the two panels, we see that the absorption feature associated with the CH$_3$OH 1(1,1)- 0(0,0)++ line (E$_u$ =16.8 K) around the A1 core is extended and have a size of 0\farcs87 $\times$ 0\farcs74 (deconvolved from the beam after 2D Gaussian fittings), reflecting its low-lying energy nature. The integrated emission for the same transition toward A2 is also affected by the absorption part of the inverse P-Cygni profile.
The absorption around A1 core for the two other higher E$_u$ methanol transitions have a compact size of 0\farcs42 $\times$ 0\farcs29.
To enhance the imaging quality, for species other than CH$_3$OH, the integrated intensity maps are made by stacking the detected transitions all together. 
Based on 2D Gaussian fitting, we find that the absorption feature for $^{13}$CH$_3$OH, CH$_3$CHO, CH$_2$DOH around the A1 core are all compact.

\subsection{Gas temperatures and column densities}

Towards IRAS~4A2, there are a large number of emission lines showing noticeably similar intensity levels at around 70~K  (without continuum subtraction) as shown in Figure~\ref{line_spectra}. 
This is particularly true for the three detected CH$_3$OH transitions, all the CH$_2$DOH transitions, and at least the three low energy CH$_3$CHO transitions, as also shown in Figure~\ref{A2_spectralfit}. 
Given that these emission features originate from different species with different excitation energies and Einstein A coefficients, their common brightness can be understood only if all these lines are optically thick and saturated. The opaque nature indicates that the molecular gas at A2 is at least as hot as 70~K at its ``surface". For $^{13}$CH$_3$OH in A2, the emissions may be optically thin, although the high E$_u$ line is possibly blended by an interloper.
Additionally, a couple of high E$_u$ CH$_3$CHO transitions do not seem to be saturated.

For these cases where we have detected multiple optically thin transitions from the same molecule, we may estimate the gas temperature using the rotation diagram method (\citealp{turner91, herbst09} and references therein). %added based on Zhi_yun li's suggestion
Following the standard radiative transfer equation we have
\begin{equation}
    I_{\nu} = J_{\nu}(T_{ex})(1 - e^{-\tau_0})  + J_{\nu}(T_{bg}) e^{-\tau_0}
\end{equation}
where $I_{\nu}$ is the observed brightness at the line frequency ${\nu}$,  $J_{\nu}$ is the source function ($=(h\nu/k)[exp(h\nu/kT)-1]^{-1}$, $T_{ex}$ and $T_{bg}$ are the gas excitation temperature at that transition and the background brightness temperature. $\tau_{0}$ is the optical depth at the line center.
Equivalently, 
\begin{equation}
    I_{\nu} = J_{\nu}(T_{ex}) + (J_{\nu}(T_{bg}) - J_{\nu}(T_{ex}))  e^{-\tau_0}
\end{equation}
The observed continuum-subtracted line brightness temperature $\Delta T_B$ can be expressed as 
\begin{eqnarray}
    \Delta T_B &=& I_{\nu} - J_{\nu}(T_{bg}) \\
                      &=& [J_{\nu}(T_{ex}) - J_{\nu}(T_{bg}](1 - e^{-{\tau_0}})
\end{eqnarray}
Equation (4) is the same as Equation (A1) of \citet{turner91}, 
When the line emission is assumed optically thin, the upper level population $N_u$ is given by Equation (A3) of \citet{turner91}, 
\begin{equation}
 % N_u=\frac{8\pi k \nu^3}{h c^3 A_{ul} n_{bf}} \int T_{mb} dv \\
 \frac{N_u}{g_u} = \frac{T_{ex}}{T_{ex}-T_{bg}}\frac{3kW}{8 \pi^3 \nu S \mu^2}
\end{equation}
where, $g_u$ is the degeneracy of the upper level, 
$k$ is the Boltzmann constant, 
S is the line strength, $\mu$ is the dipole moment,
W = $\int \Delta T_B dv$ is the integrated line intensity in K~km s$^{-1}$, and the Rayleigh-Jeans approximation has been applied.
Furthermore, the level population can be described using Boltzmann distribution, characterized by a single (rotation) temperature T$_{ex} =$ T$_{rot}$, if its excitation is assumed in local thermodynamic equilibrium (LTE).
The total molecular column density and the excitation (rotation) temperature can thus be related by the relation,
\begin{equation}
 \frac{N_u}{g_u}=\frac{N_{tot}}{Q_{tot}} e^{-E_u/K T_{rot}}
\end{equation}
where N$_{tot}$ is the total column density, Q$_{tot}$ is the rotational partition function at T$_{rot}$. 
Combining Equations (5) and (6), we can find
\begin{equation}
    log \frac{3kW}{8 \pi^3 \nu S \mu^2 } = log \frac{N_{tot}}{Q_{tot}} - \frac{E_u}{k}\frac{1}{T_{rot}} - log \frac{T_{rot}}{T_{rot} - T_{bg}}
\end{equation}
This equation, with a negligible $T_{bg}$, hence a negligible last term, forms the conventional rotation diagram. That is, for the same molecule with multiple detected transitions, one can plot log(N$_u$/g$_u$) vs. E$_u$ and determine T$_{rot}$ and N$_{tot}$.
We note that, though, when $T_{bg}$ cannot be fully neglected but the line opacity remains reasonably small ($< 1$), the plot of log(N$_u$/g$_u$) vs. E$_u$ still gives rise to a slop of 1/$k$ T$_{rot}$.

\begin{figure*}[ht]
\includegraphics[width=8cm]{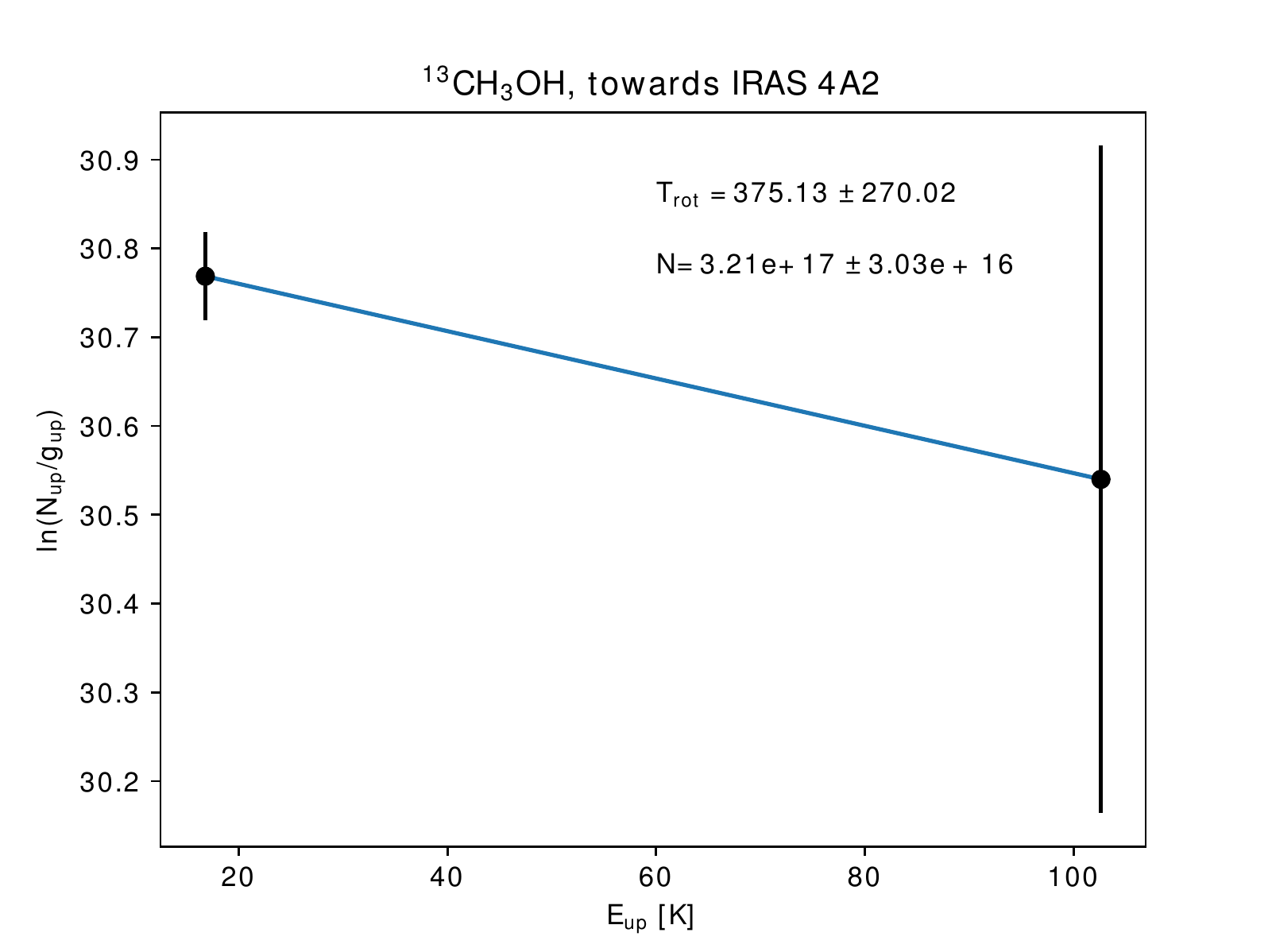}
\includegraphics[width=8cm]{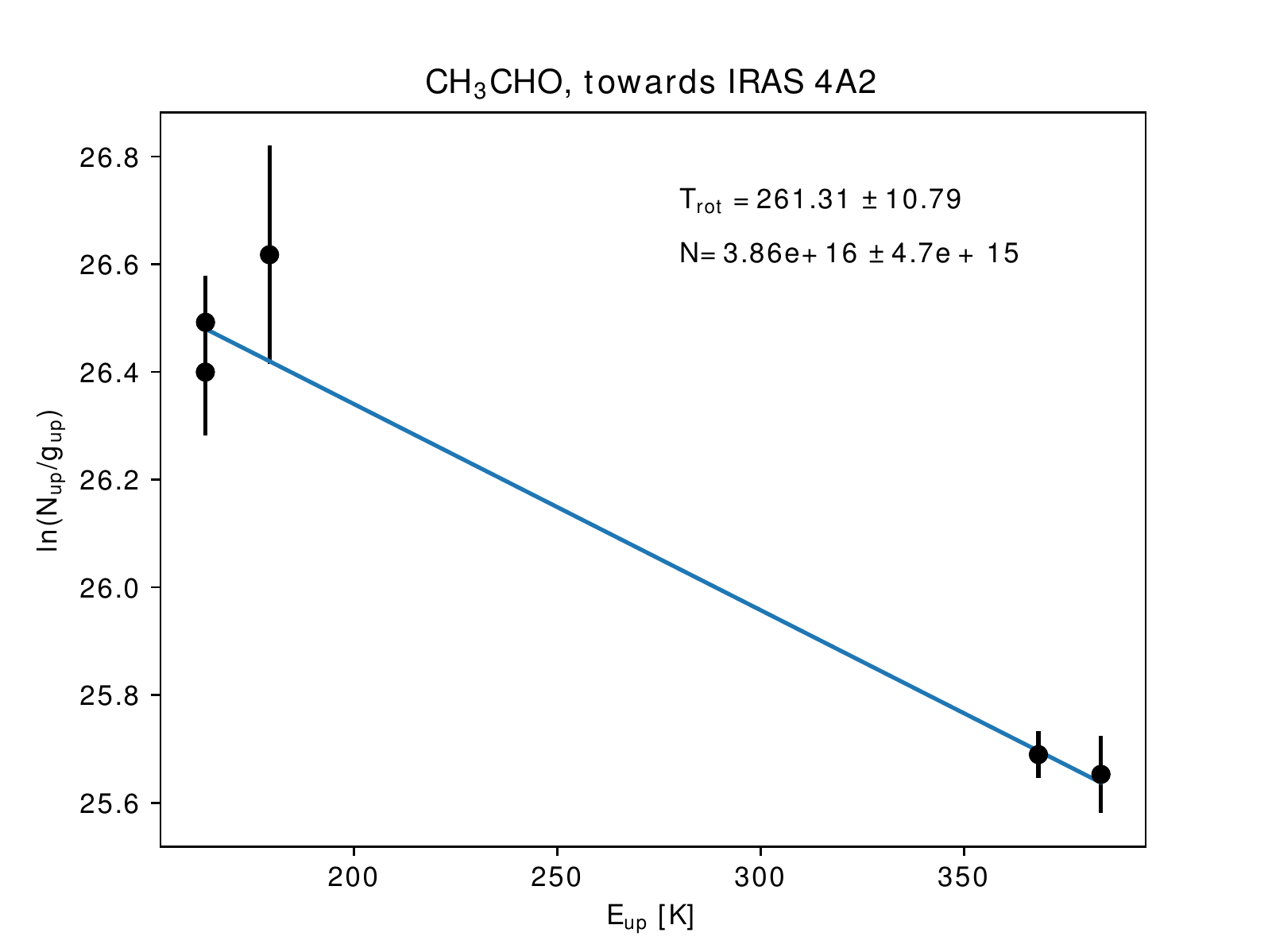}

\vskip -0.4cm
\includegraphics[width=8cm]{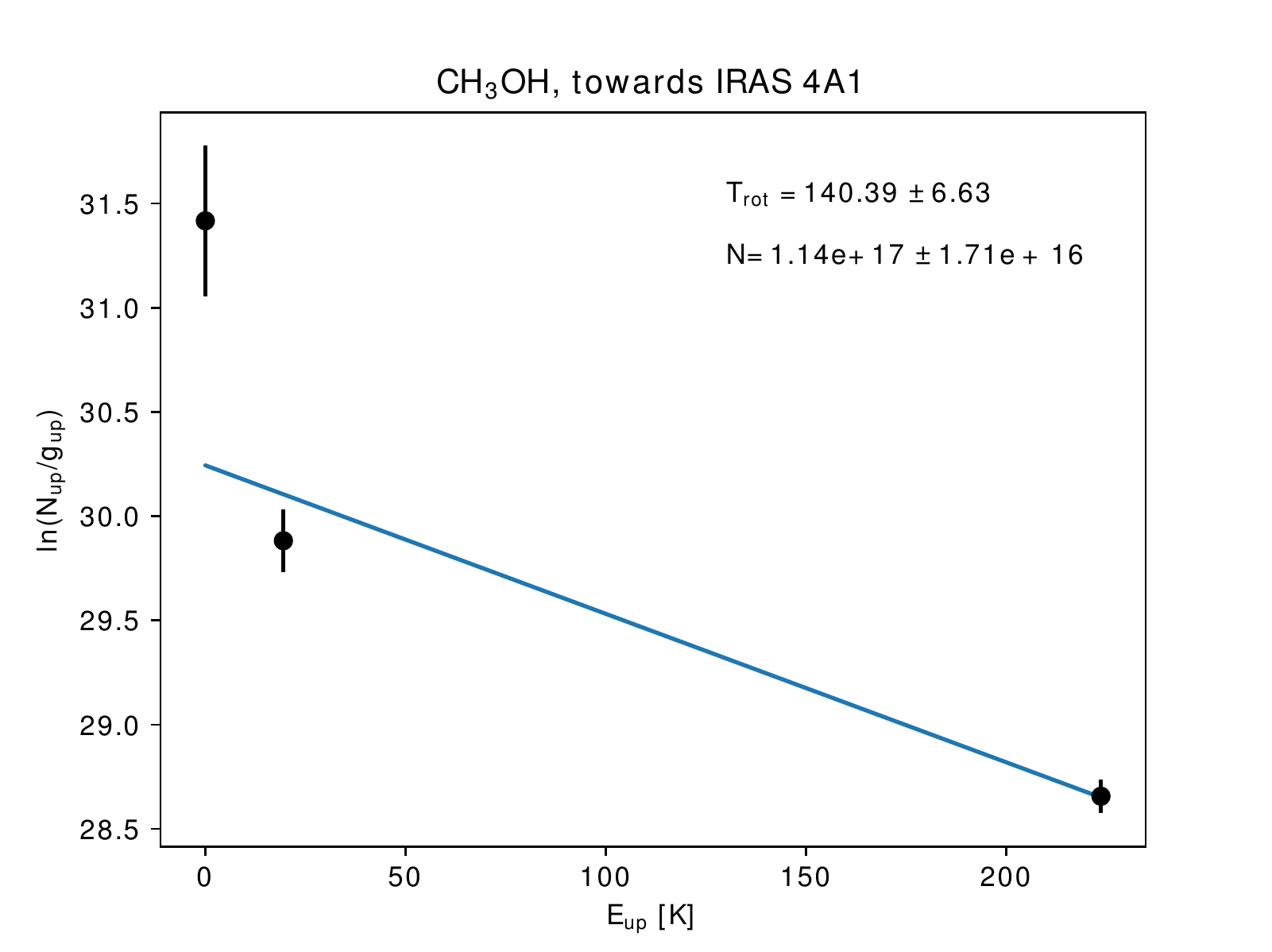}
\includegraphics[width=8cm]{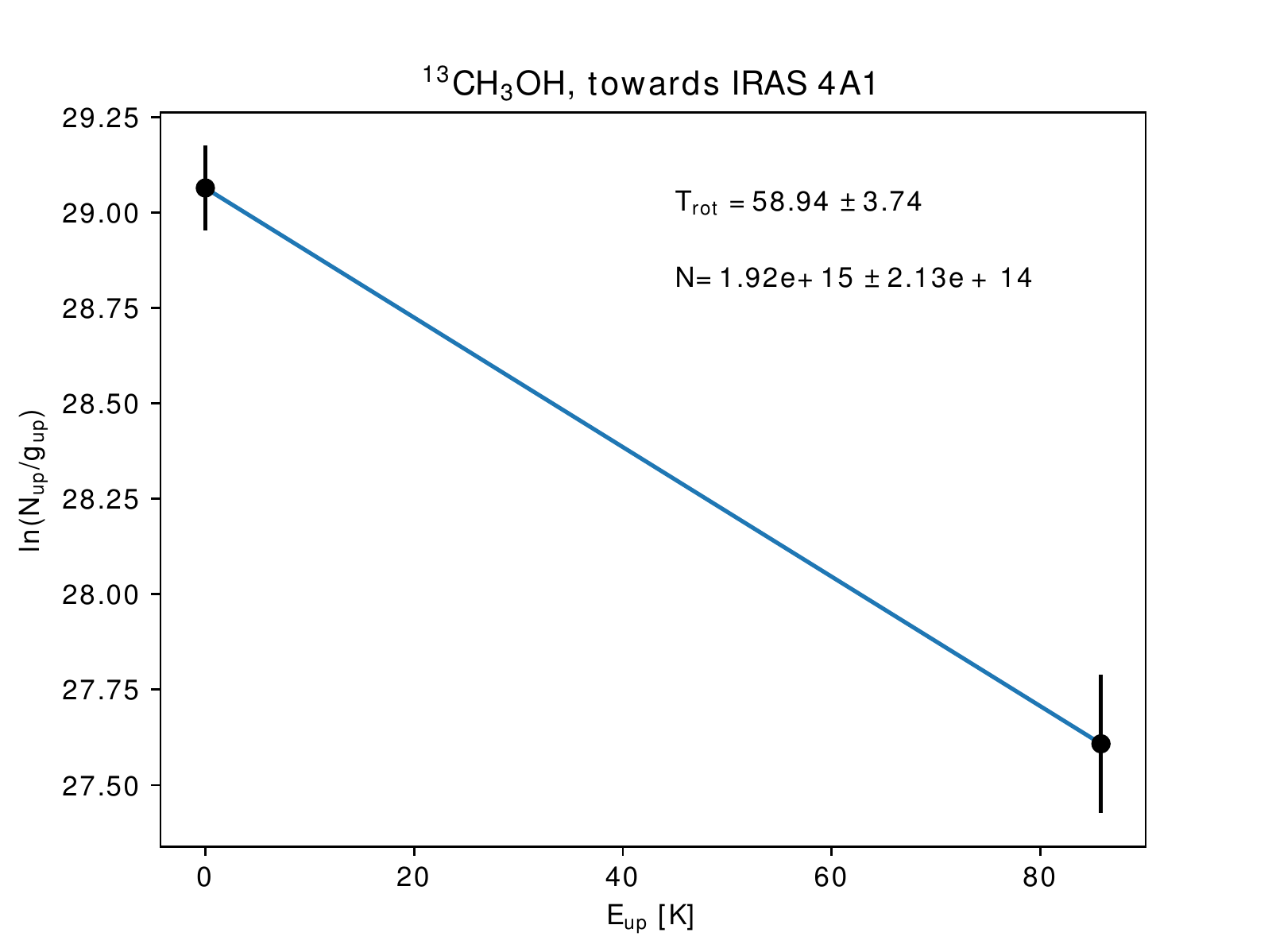}
\caption{Rotational-diagram of $^{13}$CH$_3$OH, CH$_3$CHO towards IRAS 4A2 and $^{13}$CH$_3$OH, CH$_3$OH towards IRAS 4A1. Best fit value of rotational temperature and column density is mentioned in the panels.}
\label{rot_dia}
\end{figure*}

We applied the rotation diagram method to the $^{13}$CH$_3$OH and the CH$_3$CHO transitions detected towards A2. As shown in Figure~\ref{rot_dia}, we obtained rotation temperatures of 375.13$\pm$270.02~K and 261.31$\pm$10.79~K for $^{13}$CH$_3$OH and the CH$_3$CHO, respectively. These temperatures should be taken as upper limits though, as their corresponding lower E$_u$ lines are possibly optically thick.
As \citet{goldsmith99} pointed out that, when the optical depth is not negligible, the true total column density $N_{tot}$ is related to the column density under the optically thin condition $N_{tot}^{thin}$ by the following relation,
\begin{equation}
 N_{tot}=N_{tot}^{thin} \frac{\tau}{1-exp(-\tau)}.
\end{equation}
That is, when the low E$_u$ lines are optically thick, their respective N$_u$ need to be corrected upwards, leading to a steeper slope in the rotation diagram and hence a lower rotation temperature. We also obtained lower limits on their column densities. Indirectly we can have an educated guess for the CH$_3$OH column density also by considering the isotopic ratio $\rm{^{12}C/^{13}C} \sim$~70 \citep{sheffer07}.

The situation toward IRAS~4A1, for its exclusively absorption features, can be more complicated. As the absorption features have various depths, in a naively scenario (Scenario I, see Figure~\ref{A1_diagram}), we assume that most, if not all, of the observed transitions in A1 are optically thin and there is no saturation in absorption.
We regard the continuum emission as the ``background" radiation in the radiative transfer equation and the absorbing molecular line forms in a ``foreground" layer. We would therefore have both ($T_{ex} - T_{bg}$) and $W$ terms in Equation (5) negative as $T_{ex} < T_{bg}$ and $\Delta T_B < 0$ and proceed with the rotation diagram method. 

When this Scenario I is applied to the CH$_3$OH and $^{13}$CH$_3$OH transitions in A1, we obtain Figure~\ref{rot_dia}, which indicates that the corresponding rotation temperatures for CH$_3$OH and $^{13}$CH$_3$OH are 140.39$\pm$6.63~K and 58.94$\pm$ 3.74~K, respectively.
While CH$_2$DOH does not fit with a well defined rotational temperature probably due to line blending.
The detected transitions of CH$_3$CHO towards A1 may be optically thin but they have very closely spaced E$_u$ values ($\sim 163-179$~K) and do not provide good constraints on the temperature through a rotation diagram. The column densities are shown in Table~\ref{col_den}, for molecules where do not have any estimation of rotational temperature, we choose the temperature to be 200~K and calculated the column densities averaging 
the results from all the detected transitions of a molecule. 

\begin{deluxetable*}{lllll}
\tablecaption{Average column density of the detected molecules in IRAS 4A \label{col_den}}
\tablehead{
\colhead{Molecule} & \colhead{T$_{rot}$}& \colhead{column density}&\colhead{T$_{rot}$}&\colhead{column density}}
\startdata
              & (K) & N (cm$^{-2}$) & (K) & N (cm$^{-2}$) \\
\hline
\colhead{} & \multicolumn{2}{|c|}{IRAS 4A2} & \multicolumn{2}{c|}{IRAS 4A1} \\
\hline  
 CH$_3$OH & 200 K & $\geq (5.47 \pm 0.12) \times 10^{17}$ \tablenotemark{b} &$\leq 140.39 \pm 6.63$~K & $\geq (1.14 \pm 0.17)\times 10^{17}$ \\
 $^{13}$CH$_3$OH & $\leq 375.13 \pm 270.02$ K & $\geq (3.21\pm 0.30)\times 10^{17}$ & $\leq 58.94 \pm 3.74$ K  & $\geq (1.92 \pm 0.21) \times 10^{15}$\\
 CH$_2$DOH  & 200 K $^a$ & $\geq(1.30 \pm 0.26 )\times 10^{17}$ & 200 K $^a$ & $\geq(6.51 \pm 1.83) \times 10^{15}$\\
 CH$_3$CHO & $\leq 261.31 \pm 10.79$ K  & $\geq (3.86 \pm 0.47) \times 10^{16}$ & 200 K $^a$ & $\geq (3.2 \pm 0.67) \times 10^{14}$
\enddata
\tablenotetext{a}{Rotational temperature is not available, this is the assumed value.
The errors in the table are for fitting only; there are calibration error $\sim 10\%$, we have not included it here.} \tablenotetext{b}{For multiple transitions where there is no available rotational temperature, the average column density for the assumed temperature.}
\end{deluxetable*}

\section{Discussion}

While the observed forest of emission line features of COMs supports the hot corino nature of IRAS~4A2 as was established by previous studies (e.g. \citealp{taquet15,santg15,lopez17}), the nature of A1 is much less clear.
\citet{taquet15} made detailed studies of COMs in selected low mass YSOs, including NGC~1333 IRAS4A. Even though the angular resolution of their observations only marginally resolved apart A1 and A2, it was clear that A1 is devoid of COM emission features. 
A lack of COM emission lines toward A1 was noticed by \citet{santg15} in their study of the bipolar outflows associated in IRAS4A.
Based on the velocities and extents of the bipolar outflows, \citet{santg15} proposed a younger dynamical age of A1 as compared to A2. 
Using observations with PdBI and ALMA, \citet{lopez17} further reaffirmed a significant contrast of COM features between A1 and A2.
\citet{lopez17} detected no COM emission but marginal absorption features of CH$_3$CHO, HCOOCH$_3$ and (CH$_3$)$_2$CO towards A1 and attribute these features to
either an opaque absorbing foreground or a very small severely beam-diluted emission region. In particular, they estimated that the size of the emitting region should be of order 15 AU (after scaling the adopted distance to  293 pc) or smaller  if it is not obscured by the foreground dust.

\subsection{Continuum Opacity \label{sec:dis-cont}}

In contrast to previous presumptions, we reason below several lines of evidence that points
to the A1 continuum emission from centimeter to submillimeter wavelengths being optically thick.
First of all, the rich absorption lines in combination of the absence of emission feature we observed at 0.84~mm is an indication of A1 being either genuinely deficient of trace molecular species within or the continuum is optically thick and has attenuated the embedded molecular emissions.

Secondly, as presented in Section~\ref{sec:res-dust}, the brightness temperature of the compact component towards A1 is nearly 60~K at an angular resolution of 0\farcs3 by 0\farcs2.
At 1.2~mm (250~GHz), \citet{lopez17} detected with ALMA a peak continuum flux of 542~mJy towards the A1 centroid at an angular resolution of 0\farcs66 by 0\farcs35. This leads to a brightness temperature of 46~K.
With the Karl G. Jansky Very Large Array (JVLA) 6.9~mm observations at a resolution of 0\farcs49 $\times$ 0\farcs40, \citet{liu16} also reported a fairly high brightness temperature of 41~K towards A1.
Recently  VLA observations for the VLA Nascent Disk and Multiplicity (VANDAM) survey measured peak continuum fluxes of 4.151 mJy and 2.759 mJy at 8.1 mm and 10.3 mm, respectively, towards A1 \citep{cox15}. With their  synthesized beam size of 0\farcs224 by 0\farcs199 and 0\farcs25 by 0\farcs25, these flux densities correspond to a brightness temperature of 83 K at 8.1 mm and 64 K at 10.3 mm.
Given dust opacity is more optically thin at longer wavelengths, that is, scaling as $\nu^{\beta}$ with $\beta \sim$ 1.0 or higher, one would expect at similar angular resolutions the continuum brightness temperature measured at a lower frequency will be significantly weaker than that observed at a higher frequency.
The beam dilution effect could further make the brightness lower if the low frequency observations were done at a coarser angular resolution.
We, for example, smooth our ALMA 0.84~mm continuum to have the same beam size as that of the JVLA 6.9~mm observations and find the resulting A1 peak brightness be 39 K, almost identical to the JVLA  6.9~mm measurement.
With observations of these bands considered, the above-mentioned comparable brightness from 10~mm to 0.84~mm therefore implies opaque continuum emission throughout these bands towards the A1 peak, and the dusty opaque zone could be as large as the JVLA  beam area of 0\farcs49 $\times$ 0\farcs40. In fact, a possible opaque nature of the 1.2~mm continuum emission of A1 was considered by \citet{lopez17}, although it was ruled unlikely by the authors for its corresponding large volume density of $10^{11}$ cm$^{-3}$ towards the region.
Meanwhile, in a spectral energy distribution (SED) modelling and analysis, \citet{li17} considered a two-component (disk+envelope) fitting, which results in a good representation of the measured SED data points and hinted at A1 continuum being optically thick even down to centimeter wavelengths.

The possibility of A1 being optically thick throughout the centimeter to submillimeter bands has notable implications. All the column densities and masses reported by previous studies should then only be considered as lower limits, as they were derived with the optically-thin assumption. The lack of emission spectral feature and hence the deficiency in trace (COM) species, compared to that of A2 shall also be re-considered, as their emission could be naturally blanked by the optically thick continuum.
Additionally, the weak NH$_3$ line emission observed by \citet{choi07, choi11} towards A1 could also be partially understood if the continuum opacity remains substantial at 1.3~cm.

In the case of A2, \citet{liu16} reported a much weaker brightness of $\sim$2~K at 6.9~mm, about a factor of 20 lower that our ALMA 0.84~mm results. Note that the 0.84~mm brightness temperature smoothed to the JVLA 6.9~mm beam size is still as high as 23~K.  The brightness difference indicates that the 6.9~mm continuum towards A2 peak is likely optically thin.

\subsection{COM Features \label{sec:dis-line}}

We have detected spectra features of CH$_3$OH, CH$_2$DOH, $^{13}$CH$_3$OH and CH$_3$CHO in both A1 and A2 of IRAS 4A.
Towards A2, all features from these species, except CH$_3$OH 1(1,1)-0(0,0)++, are in emission. Many of the transitions are, as demonstrated in Section~\ref{sec:res-line}, optically thick with a brightness temperature of 70K suggesting that the gas is warm. 
The CH$_3$OH 1(1,1)-0(0,0)++ line shows an inverse P-Cygni profile, indicative of gas infall motion, and this will be discussed in a separate paper.
Towards A1, our observations demonstrate conclusively, for the first time, absorption features from the selected COMs against the optically thick dust continuum emission.

Could both the absorption features toward A1 and the emission features toward A2 have a common, possibly foreground, origin?
\citet{yildiz13} suggested tentatively a foreground layer of gas based on O$_2$ observations with the \textit{Herschel Space Observatory} toward IRAS~4A and this component has been incorporated for optimizing the model fitting of H$_2$O spectra observed also with Herschel by \citet{mottram13} in IRAS~4A.
This foreground cloud with an estimated velocity at around 8.0 \kms also seems present in the $^{13}$CO (2-1) spectrum reported by \citet{jorgen07} (their Fig.~6) in the PROSAC survey. However, this velocity component is 0.8 -- 1.0 \kms offset from the systemic velocity of IRAS~4A and incompatible with our observed COM velocities.
In addition, the spectral features of these COMs along the two close line-of-sights (towards A1 and A2) have disparate velocities, line-widths, and optical depths, strongly ruling out a common (foreground) origin of the observed COM absorption and emission.

Given that COMs have been observed sometimes in shocks associated with outflows \citep{sugimura11}, there is the possibility that the COM features are associated with the outflows in IRAS~4A.
\citet{santg15} found a high-velocity jet from A1, while A2 drives a slower precessing jet, with both outflows primarily aligned along the north-south direction.
On the other hand, our observed absorption and emission COM features rather trace closely the bright dust continuum.
Furthermore, the observed line-widths of the COM features toward the A1 and A2 continuum are relatively narrow, in contrast to the high outflow velocities several tens of \kms relative to the ambient in this region \citep{santg15}. 
An outflow origin of the COM features towards A1 and A2 is therefore unlikely, leaving the observed COMs most probably associated directly with the individual A1 and A2 continuum sources at $\lesssim$ 0.3\farcs or $\lesssim$  88 AU scale.

As shown in Figure~\ref{spectral_maps}, the low E$_u$ ($\sim16$ K) CH$_3$OH line exhibits absorption in a region that covers the full extent of the A1 continuum emission with a size of 0.9$''$. This extended absorption layer is likely associated with the protostellar envelope surrounding the central protostar and possibly its circumstellar disk. %changed based on Zhi-Yun suggestion
On the other hand, other COM absorption features from higher upper energy levels ($\gtrsim$ 40~K) arise from a barely resolved region $\le 0.25''$ or $\lesssim$  73~AU in diameter.

\begin{figure*}[ht]
\centering
\includegraphics[width=13.5cm]{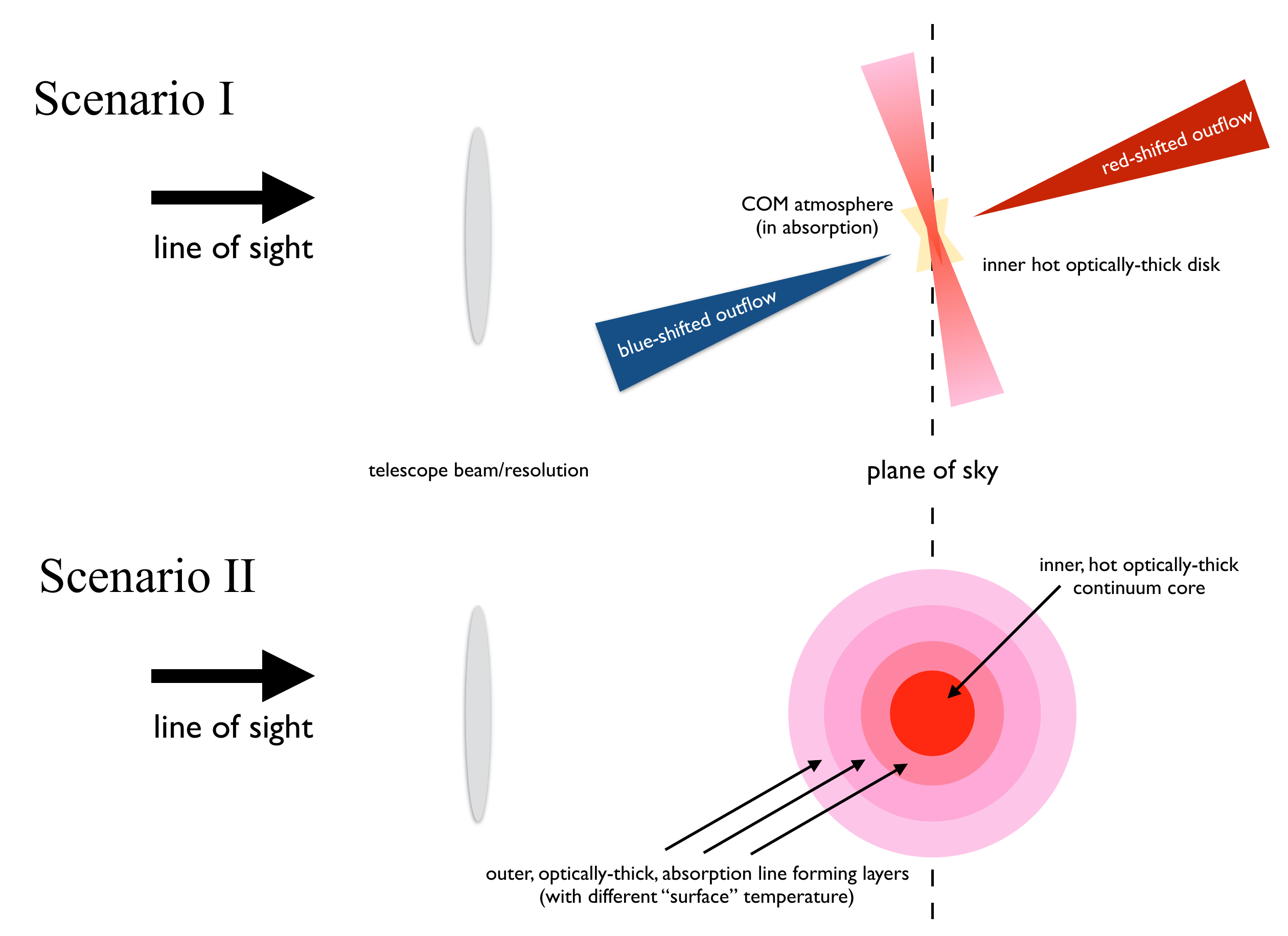}
\caption{Schematic diagram for NGC1333 IRAS4A1 configuration.}
\label{A1_diagram}
\end{figure*}

Based on the rotation diagram analysis in Scenario I, the gas temperature of this more compact absorption layer is $\sim$ 100~K.
To have the observed absorption feature produced by a layer of COM (e.g. CH$_3$OH gas), a brighter (background) continuum emission is required.
Typically, the disk upper layers are hotter due to stellar irradiation, while the midplane is cooler. On the other hand, high accretion rate can potentially heat up the midplane region, particularly at small radii, and make it hotter than its atmosphere \citep{dullemond07}.
As illustrated in Figure~\ref{A1_diagram}, in this scenario I, we speculate that we are witnessing a compact path of COM atmosphere absorbing against the very inner and hot part of an inclined circumstellar disk around a protostar at A1. 
We find, as listed in Table~\ref{col_den}, the column density of CH$_3$OH at the level of 10$^{17}$ cm$^{-2}$ (converting $^{13}$CH$_3$OH column density). Factoring in a typical fractional abundance of CH$_3$OH of 10$^{-6}$ \citep{herbst09}, the absorbing COM atmosphere will have a column density of 10$^{23}$ cm$^{-2}$. This can be compared to the column density of the ``background" continuum disk of 10$^{26}$ cm$^{-2}$ (see section~3.1).
If the disk is only slightly inclined from face-on, we are able to observe the unresolved, compact ($<$ 73~AU), hot inner disk and the associated COM atmosphere, which dominates the absorption spectral along its line of sight. 
The tentative COM absorption features seen by \citet{lopez17} may originate from the same region but get severely beam-diluted at a lower angular resolution.
The (optically thick) disk surface has a temperature gradient, which results in an overall (beam averaged) continuum brightness of 60~K towards the A1 peak. 
In the case that the level of accretion heating in A1 is not sufficient to elevate the disk mid-plane temperature above its COM atmosphere locally (i.e., at the same radius), we may be viewing a tilted disk so that along the line of sight, we are seeing an atmosphere against the continuum emission at a smaller radius with higher temperature.
\citet{seguracox18}, for example, based on two-dimensional image fitting of the 8mm (dust) continuum emission observed with the JVLA, inferred a modestly inclined disk at 35$^{\circ}$.
The slightly inclined disk probably leads to a seemingly faster but less extended bipolar outflow in A1 as compared to that in A2 \citep{santg15}, as the outflow will be highly inclined, resulting in a larger velocity along the light-of-sight and less transverse motion on the plane of the sky. 
A nearly pole-on (6$^{\circ}$) view of a bipolar outflow toward A1 was actually considered by \citet{santg15} although a different interpretation was preferred. 
A nearly face-on disk orientation could explain the lack of a clear velocity gradient seen in NH$_3$ toward A1 by \citet{choi07,choi11}.
In contrast, a velocity gradient perpendicular to the bipolar outflow was seen in NH$_3$ towards A2.  
The lack of obvious fast rotating signature in NH$_3$ emission around A1 led \citet{choi11} to speculate that A1 hosts a less massive protostar than that of the A2. 

The geometry of IRAS 4A1 proposed above is close to the picture suggested by \citet{oya18} of I16293~B.
I16293 is another proto-binary where two (sub)millimeter continuum emission cores, I16293~A and B, show hot-corino activities \citep{oya16}.
Similar to the case of IRAS~4A, the weaker continuum of I16293 (i.e., A core) show richer COM (emission) features compared to that of B core. 
\citet{zapata13} revealed a compact structure from continuum emission (0.45mm, angular resolution $\sim ~0.2''$) of I16293~B, and revealed a pronounced inner depression or absorption hole from
H$^{13}$CN, HC$^{15}$N, CH$_3$OH images. 
The `hole' size is comparable to continuum size which is optically thick. So, in that image the molecular emission is surrounded like a wall. Our observation has a little similarity with this case but in our case we do not find any molecular emission outside of the central beam.
The outer disk (beyond 36~AU in radius) is probably cold and/or optically thick so that COM emission is absent due to a lack of COMs in the gas phase and/or the attenuation of line emissions by continuum. A relevant case related to this picture is the COMs environment observed in HH212
system \citep{lee17b}. COMs were observed above the disk surface but not in the optically thick disk \citep{lee17a} itself in HH212. 

It is interesting to note that emission of complex organics such as CH$_3$OH, CH$_2$DOH, HCOOCH$_3$, and NH$_2$CHO have been proposed to trace the centrifugal barrier of accretion disks \citep{lee17b, oya18}.
In the case of HH212, \cite{lee17b} observed a couple of layers of hot COM gas, including CH$_3$OH, CH$_2$DOH, and NH$_2$CHO, above and below of an edge-on disk. %modified
The extent of the COM ``atmosphere" is consistent with that of a centrifugal barrier (CB) at 44~AU based on kinematics considerations.
For I16293~B, compact CH$_3$OH and HCOOCH$_3$ emission were observed toward its continuum; it was also suggested to trace the CB at a radius of 40~AU.
If our observed COM emission is indeed associated with the same phenomenon, the compact nature of the COM absorption in A1 would imply that only a small ($\sim$ 36 AU) disk within which Keplerian rotation has been established.
We have not seen all the molecular transitions that are observed towards A2 in the absorption towards A1 (see Figure~\ref{line_spectra}). 
The difference perhaps indicates that COM emission from A2 has a ``genuine"/``typical" hot-corino nature while that from A1 is accretion-shock related at CB.

In addition to Scenario I in which the absorption lines are considered optically-thin, we discuss also an alternative scenario (Scenario II). It is possible that some (or all) of the absorption features toward A1 are in fact optically thick and saturated. As shown in Equation (2) and (4), in the optically thick situation, the depth of the absorption line core will be the temperature difference between the background continuum brightness and the foreground absorbing gas temperature, or equivalently, the brightness at the absorption line core is equal to the foreground absorbing gas temperature. 
In Scenario II, in which invoking a beam-filling factor is not necessary,
the absorption lines with different excitation temperatures form at different layers in a temperature-stratified envelope, a configuration illustrated in Figure~\ref{A1_diagram}. The lower excitation line saturates at the outer cooler layer, while the higher excitation transition (not excited at the outer layer) become optically thick in the relatively inner and hotter region. 
The CH$_3$OH transitions toward A1, for example, probe layers between 30K and 60K.
Considering an envelope with its gas temperature scaling with r$^{-0.5}$,
where r is the radius from the center protostellar object, a factor of two difference in temperature translates to a radius difference by a factor of 4. 
The velocity for gas, if free-falling at 50 AU around a 0.1M$_\odot$ protostar, is nearly 2 km~s$^{-1}$ and would be scaled by $r^{-0.5}$. 
Our observations should be able to detect such (differential) velocity offsets if the gas is under free-fall.
As suggested in Section 3.2, the consistency of the absorption line velocities with the systematic velocity and among themselves appears to indicate a lack of gas accretion/inflow motion (along the line of sight) in the envelope at tens of AU scale.

\subsection{Deuterium Fractionation \label{sec:dis-deuterium}}

Now we consider the deuterium fractionation of CH$_3$OH in A1 and A2. As CH$_3$OH emission in A2 is optically thick, it would be good to compare $^{13}$CH$_3$OH and CH$_2$DOH column densities in both the cores. The $^{13}$CH$_3$OH/CH$_2$DOH ratio for A2 and A1 is 2.47 and 0.29 (considering the average value) respectively. \citet{bainchi17} used $^{13}$CH$_3$OH to calculate $^{12}$CH$_3$OH abundance and from this they reported a D/H ratio $\simeq 2\times 10^{-2}$ for methanol in HH212. 
The deuterium fractionation is one order lower than typical D/H values in hot-cores and prestellar clouds. The observation by \citet{bainchi17} was a high-resolution ALMA observation on scale 45 AU, similar to the scale of current observation $\sim$73 AU.
If we consider $^{12}$C/$^{13}$C $\sim$ 70 \citep{sheffer07} (note: $^{12}$C/ $^{13}$C ratio is not standard and it varies in different medium, e.g., \citealp{tani16,wirs11}), then the lower limit of CH$_2$DOH/CH$_3$OH ratio in A2 and A1 becomes the 0.6$\times 10^{-2}$ and 4.9$\times 10^{-2}$ respectively. This is only a factor $\sim 8$ difference in deuteration between the core A1 and A2. There is a caveat in this calculation,
we assume a likely value of rotational temperature (200~K) for this and also there is huge uncertainty in $^{13}$CH$_3$OH column density calculation, and rotational temperatures of $^{13}$CH$_3$OH for two cores are very different. However, if we consider 
the result, then it suggests that higher gas temperature in the hot-corinos environment indirectly reduces 
the deuteration in methanol compared to cold prestellar conditions; this is similar to the methanol deuteration in HH212. It may imply that A1 also have a hot environment like the A2 with similar deuterium fractionation and hosts a hot-corino.

\section{Conclusion}

We have observed the NGC1333 IRAS 4A protobinary system with its two cores- A1 \& A2  well resolved using ALMA. The results can be summarized as follows-

1. The dust continuum emission towards the core A2 is found to be predominantly optically thin at this scale while the continuum emission towards A1 is most likely to be optically thick.

2. A forest of spectral line emissions is observed towards A2, while spectral transitions towards A1 are detected in absorption. Here we have identified and discussed some COMs like CH$_3$OH,  $^{13}$CH$_3$OH,  CH$_2$DOH, CH$_3$CHO in both the cores A1 and A2.  The A2 core is a known hot corino with a rich presence of COMs.

3. The observed absorption features towards A1 are probably arising from a hot-corino-like atmosphere against a very compact ($\leq$ 36 AU) disk in A1. We speculate that this compact hot-corino-like atmosphere may resemble the cases in HH212 and I16293~B in which the COM emission is related to the centrifugal barrier of accretion disks. Alternatively, the absorption may arise from different layers of a temperature-stratified dense envelope. 

4. An indirect calculation of deuterium fractionation shows that CH$_2$ DOH/CH$_3$OH ratio have similar order ($\sim 10^{-2}$) towards both the sources. This low deuterium fractionation in both sources may imply a hot gas condition ($\sim 100$~K) typical to a hot corino.

 {\bf Acknowledgment}
 We thank Prof. Emmanuel Caux for helpful discussion and suggestions, Dr. Kuo-Song Wang, Dr. Tien-Hao Hsieh (ASIAA, Taiwan) for some technical help.  DS wants to thank ASIAA, Taiwan for providing a visitor position and computational facility; also wishes to thank PRL, India for supporting the research with post-doctoral fellowship and international travel grant. SYL acknowledges the support by the Minister of Science and Technology of Taiwan (MOST 107-2119-M-001-041). ZYL is supported in part by NSF grant AST-1815784 and AST-1716259 and NASA grant 80NSSC18K1095 and NNX14AB38G. 
 ST acknowledges a grant from the Ministry of Science and Technology (MOST) of Taiwan (MOST 102-2119- M-001-012-MY3), and JSPS KAKENHI Grant Numbers JP16H07086 and JP18K03703 in support of this work. This work was supported by NAOJ ALMA Scientific Research Grant Numbers 2017-04A.
 This paper makes use of the following ALMA data: ADS/JAO.ALMA\#2015.1.00147.S. ALMA is a partnership of ESO (representing its member states), NSF (USA) and NINS (Japan), together with NRC (Canada) and MoST and ASIAA (Taiwan) and KASI (Republic of Korea), in cooperation with the Republic of Chile. The Joint ALMA Observatory is operated by ESO, AUI/NRAO and NAOJ.

\end{document}